\documentclass[12pt,a4paper]{article}

\usepackage{fullpage}
\usepackage{cite}
\usepackage{graphicx}
\usepackage{jheppub}
\usepackage{subfigure}
\date{}

\def\eq#1{Eq.~(\ref{#1})}

\newcommand{\secn}[1]{Section~\ref{#1}}

\newcommand{\be}{\begin{equation}}
\newcommand{\ee}{\end{equation}}
\newcommand{\bea}{\begin{eqnarray}}
\newcommand{\eea}{\end{eqnarray}}
\newcommand{\nn}{\nonumber}

\newcommand{\as}{\alpha_s}
\newcommand{\eps}{\epsilon}
\newcommand{\sig}{\sigma}
\newcommand{\e}{\epsilon}
\newcommand{\beq}{\begin{eqnarray}}
\newcommand{\eeq}{\end{eqnarray}}
\newcommand{\npo}{{n+1}}
\newcommand{\npt}{{n+2}}
\newcommand{\pn}{\Phi_n}
\newcommand{\pnpo}{\Phi_\npo}
\newcommand{\pnpt}{\Phi_\npt}

\newcommand{\LO}{{\mbox{\tiny{LO}}}}
\newcommand{\NLO}{{\mbox{\tiny{NLO}}}}
\newcommand{\NNLO}{{\mbox{\tiny{NNLO}}}}
\newcommand{\CG}{{\mbox{\tiny{CG}}}}
\newcommand{\E}{{\mbox{\tiny{E}}}}

\newcommand{\one}{\, (\mathbf{1})}
\newcommand{\two}{}
\newcommand{\otwo}{\, (\mathbf{12})}
\newcommand{\ttwo}{\, (\mathbf{2})}
\newcommand{\RV}{\, (\mathbf{RV})}
\def\eq#1{Eq.~(\ref{#1})}
\def\bra#1{%
  \left\langle\smash{#1}{\vphantom1}\right|}
\def\ket#1{%
  \left|\smash{#1}{\vphantom1}\right\rangle}
\def\slash#1{#1 \hskip-0.45em /}

\def\slash#1{#1 \hskip-0.45em /}

\title{Factorisation and Subtraction beyond NLO}

\author{L. Magnea,}
\author{E. Maina,}
\author{G. Pelliccioli,}
\author{C. Signorile-Signorile,}
\author{P. Torrielli and}  
\author{S. Uccirati.}

\vspace*{1cm} 

\affiliation{Dipartimento di Fisica and Arnold-Regge Center, Universit\`a di Torino,\\ 
and INFN, Sezione di Torino, Via P. Giuria 1, I-10125 Torino, Italy}

\emailAdd{magnea@to.infn.it}
\emailAdd{maina@to.infn.it}
\emailAdd{gpellicc@to.infn.it}
\emailAdd{signoril@to.infn.it}
\emailAdd{torriell@to.infn.it}
\emailAdd{uccirati@to.infn.it}

\abstract{We provide a general method to construct local infrared subtraction
counterterms for unresolved radiative contributions to differential cross sections, 
to any order in perturbation theory. We start from the factorised structure of virtual 
corrections to scattering amplitudes, where soft and collinear divergences are 
organised in gauge-invariant matrix elements of fields and Wilson lines, and 
we define radiative eikonal form factors and jet functions which are fully 
differential in the radiation phase space, and can be shown to cancel virtual 
poles upon integration by using completeness relations and general theorems
on the cancellation of infrared singularities. Our method reproduces known 
results at NLO and NNLO, and yields substantial simplifications in the organisation
of the subtraction procedure, which will help in the construction of efficient 
subtraction algorithms at higher orders.}

\keywords{Factorisation, Subtraction, Perturbative QCD, Jets, Wilson lines.}

\begin{document}
\maketitle
\allowdisplaybreaks 

%%%%%%%%%%%%%%%%%%%%%%%%%%%%%%%%%%%%%%%

\section{Introduction}
\label{Introd}

Infrared divergences arising from exchanges of soft and collinear massless 
particles are well known to cancel in infrared-safe observable cross sections,
where singularities in virtual corrections to scattering amplitudes are compensated
by divergences arising from the phase-space integration of unresolved real 
radiation~\cite{Bloch:1937pw,Kinoshita:1962ur,Lee:1964is,Grammer:1973db}.
The concrete implementation of this cancellation in perturbative calculations
for massless gauge theories is relatively straightforward for low-multiplicity
final states and for highly inclusive cross sections, where the involved phase-space
integrals and the structure of typical observables are sufficiently simple (witness, 
for example, the four-loop calculation of the total cross section for
annihilation of electroweak gauge bosons into hadrons~\cite{Baikov:2008jh,Baikov:2012er}).
The situation is considerably more challenging for higher multiplicities
and for typical collider observables, where real radiation is subject to intricate
phase-space constraints, possibly involving non-trivial recursive jet algorithms.
In these cases the phase-space integration must be performed numerically, 
and the cancellation of soft and collinear divergences is much more difficult 
to implement. Common approaches involve the definition of approximate 
real-radiation matrix elements with the correct singularity structure, which 
are then integrated analytically in order to achieve the required singularity 
cancellation before numerical tools are employed.

Any solution to the subtraction problem hinges upon our general understanding
of infrared divergences in perturbation theory. In particular, the structure of soft 
and collinear singularities in virtual corrections to scattering amplitudes is very 
precisely understood~\cite{Sen:1982bt,Collins:1989bt,Sterman:1995fz,Catani:1998bh,
Sterman:2002qn,Dixon:2008gr,Gardi:2009qi,Gardi:2009zv,Becher:2009cu,
Becher:2009qa,Feige:2014wja}: divergent contributions to generic massless
gauge theory amplitudes can be factorised from the hard scattering in terms
of a small set of universal functions, defined by gauge-invariant operator 
matrix elements. Furthermore, these functions obey evolution equations that
can be solved in terms of soft and collinear anomalous dimensions, which
are completely known in the massless case up to three loops~\cite{Almelid:2015jia,
Almelid:2017qju}. Differential information on real-radiation matrix elements
is somewhat less detailed: the latter have been shown, in considerable generality, 
to factorise in soft and collinear limits into products of lower-point amplitudes multiplied 
times universal kernels~\cite{Kosower:1999xi,Catani:1998nv,Catani:1999ss};
all the relevant kernels needed for NNLO calculations are known~\cite{Catani:1999ss,
Campbell:1997hg,Bern:1999ry,Catani:2000pi}, with partial information available at 
N$^3$LO as well~\cite{DelDuca:1999iql, Duhr:2013msa,Li:2013lsa,Banerjee:2018ozf,
Bruser:2018rad}.

At NLO, such factorisation properties were first employed for the general cancellation
of infrared singularities in the so-called `slicing' approaches~\cite{Giele:1993dj,
Giele:1994xd}: these involve isolating singular regions of phase space by means 
of a small resolution scale (the `slicing parameter'), approximating real radiation 
matrix elements by the relevant infrared kernels below that scale, and integrating 
the latter in $d$ dimensions, so as to explicitly cancel the infrared poles of virtual 
origin. This procedure yields a correct result up to powers of the slicing parameter, 
which then has to be taken as small as possible, compatibly with numerical 
stability. In order to avoid this parameter dependence, `subtraction' algorithms
\cite{Frixione:1995ms,Catani:1996vz,Nagy:2003qn}, were later developed at 
NLO: in these schemes, one introduces local infrared counterterms containing
the leading singular behaviour of the radiative amplitudes in all relevant regions 
of phase space. One then subtracts the local counterterms from the radiative 
amplitude, leaving behind an integrable remainder, and one adds back to the 
virtual correction the exact integral of the local counterterms over the radiation 
phase space, cancelling explicitly the virtual infrared singularities; the resulting 
finite cross section can safely be integrated numerically, and the whole procedure 
is exact, not involving any approximation. These NLO subtraction algorithms are 
currently implemented in efficient generators~\cite{Campbell:1999ah,Gleisberg:2007md,
Frederix:2008hu,Czakon:2009ss,Frederix:2009yq,Hasegawa:2009tx,Alioli:2010xd,
Platzer:2011bc,Reuter:2016qbi}, and the handling of infrared singularities is not a 
bottleneck for phenomenological predictions at this accuracy.

At NNLO and beyond, the construction of general subtraction algorithms is the 
subject of intense current research. The technical difficulties are significant, due 
to the proliferation of overlapping singular regions when the number of unresolved 
particles is allowed to grow, and due to the increasing complexity of the soft and 
collinear splitting kernels at higher orders. Several schemes have been proposed 
to address the NNLO problem, belonging either to the slicing~\cite{Catani:2007vq,
Grazzini:2017mhc,Boughezal:2015dra,Gaunt:2015pea,Boughezal:2015ded,
Boughezal:2016wmq,Boughezal:2016zws,Boughezal:2018mvf} or to the subtraction 
\cite{GehrmannDeRidder:2008ug,Weinzierl:2008iv,Ridder:2015dxa,Currie:2016ytq,
Currie:2017eqf,Czakon:2010td,Czakon:2015owf,Czakon:2016ckf,Caola:2017dug,
Caola:2017xuq,Caola:2018pxp,Somogyi:2006da,Somogyi:2006db,DelDuca:2016csb,
Cacciari:2015jma} families. Novel ideas are also being introduced~\cite{Sborlini:2016hat,
Herzog:2018ily}, and the first studies of simple N$^3$LO processes have recently 
appeared~\cite{Dulat:2017prg,Currie:2018fgr,Cieri:2018oms}. The variety of NNLO 
methods developed so far underscores both the phenomenological interest and the 
technical difficulty of the problem, which so far has not been solved in full generality.
It is clear that in the near future it will become phenomenologically relevant,
and theoretically interesting, to extend the application of NNLO methods to more
complicated processes, and to devise subtraction algorithms at higher orders.
Such extensions will require a high degree of optimisation of existing procedures,
and possibly the implementation of new methods and theoretical ideas.

In this paper, we propose a theoretical framework to systematically analyse
the structure of soft and collinear local subtraction counterterms to any order in 
perturbation theory. Our guiding principle is the well-understood structure
of infrared divergences in virtual corrections to scattering amplitudes. We 
note that the detailed structure of virtual factorisation must be reflected in
the organisation of local counterterms: this implies significant simplifications,
in particular for overlapping soft and collinear singularities, which are 
straightforwardly handled in the virtual case. Furthermore, we note that 
explicit high-order calculations of soft anomalous dimensions have shown
that many kinematic and colour structures which could potentially contribute
to infrared divergences are in fact absent or highly constrained, a feature
that must also be reflected in the form of the real-radiation counterterms.
Finally, we note that virtual corrections to infrared singularities exponentiate
non-trivially, providing connections between low-order and high-order 
contributions. These interesting and well-understood properties have not 
so far been fully exploited for the analysis of real-radiation subtraction 
counterterms, and we hope that our discussion in this paper will lead to 
progress in this direction. Indeed, our central result is a set of definitions 
for local soft and collinear counterterms, written in terms of gauge-invariant
matrix elements of fields and Wilson lines, and valid to all orders in perturbation 
theory, which can be shown to cancel all virtual and mixed real-virtual 
singularities on the basis of general cancellation theorems~\cite{Kinoshita:1962ur,
Lee:1964is}, and of simple completeness relations. These definitions can easily 
be shown to reproduce known results at NLO and NNLO, and provide the basis 
for a first-principle calculation of higher-order universal infrared kernels. Applying 
this technology at NNLO, we find a simple and physically transparent organisation 
of soft and collinear subtractions, including in particular the treatment of double 
counting of the soft-collinear  regions.

The paper is organised as follows: in \secn{Factor}, we briefly review the infrared
factorisation of multi-parton scattering amplitudes for massless gauge theories; 
then, in \secn{Subtra}, we present a basic outline of the subtraction problem 
at NLO and NNLO: a companion paper~\cite{Magnea:2018hab} is devoted to a
detailed construction of a full subtraction algorithm for final-state singularities;  
in Sections~\ref{Softct} and \ref{Collct}, we present our definitions for soft 
and collinear local counterterms, valid to all to all orders in perturbation 
theory; in \secn{NLOres}, we briefly illustrate the definitions by showing 
how they reconstruct the well-understood structure of final-state infrared
subtraction at NLO; in \secn{NNLOres} we apply our general results 
to the problem of NNLO subtraction, and we provide precise expressions
for all the local counterterms required for hadronic massless final states;
finally, we discuss future developments in \secn{Conclu}.

%%%%%%%%%%%%%%%%%%%%%%%%%%%%%%%%%%%%%%%

\section{Infrared factorisation for virtual corrections}
\label{Factor}

We begin by describing the simple multiplicative structure of infrared poles
that emerges from the factorisation of fixed-angle multi-particle gauge-theory 
amplitudes, in order to illustrate the potential simplification that might follow
for real soft and collinear radiation. Infrared singularities in these amplitudes 
factorise in a way which is reminiscent of the renormalisation of ultraviolet 
divergences: for an amplitude involving $n$ massless particles with momenta 
$p_i$, the result takes the form~\cite{Becher:2009cu,Becher:2009qa,
Gardi:2009zv}
\beq 
  {\cal A}_n \left( \frac{p_i}{\mu}, \as(\mu^2), \e \right) \, = \, 
  {\cal Z}_n \left( \frac{p_i}{\mu}, \as(\mu^2), \e  \right) 
  {\cal F}_n \left( \frac{p_i}{\mu}, \as(\mu^2), \e  \right) \, .
\label{IRfact}
\eeq
In this compact notation, the amplitude ${\cal A}_n$ and the finite coefficient 
function ${\cal F}_n$ are vectors in the finite-dimensional space of colour 
configurations, and the divergent factor ${\cal Z}_n$ is a colour operator.
Soft-collinear factorisation implies evolution equations, which lead to the 
exponentiation of infrared poles in terms of a finite infrared anomalous 
dimension matrix $\Gamma_n$. One may write
\beq
  {\cal Z}_n \left(\frac{p_i}{\mu}, \as (\mu^2), \e \right) \, = \,  
  {\cal P} \exp \left[ \frac{1}{2} \int_0^{\mu^2} \frac{d \lambda^2}{\lambda^2} \, \,
  \Gamma_n \left(\frac{p_i}{\lambda}, \as \! \left( \lambda^2, \eps \right) \right) \right] \, ,
\label{RGsol}
\eeq
where all infrared singularities are generated by the integration of the 
$d$-dimensional running coupling over the scale $\lambda$, extending
to $\lambda = 0$~\cite{Magnea:1990zb}. The infrared anomalous dimension 
matrix $\Gamma_n$ is strongly constrained in the massless case by the
factorisation of soft and collinear poles (see \eq{factoramp} below). In 
full generality, one writes
\beq
  \Gamma_n \left(\frac{p_i}{\mu}, \as(\mu^2) \right) \, = \, 
  \Gamma_n^{\rm dip} \left(\frac{s_{ij}}{\mu^2}, \alpha_s(\mu^2) \right) \, + \, 
  \Delta_n \left( \rho_{i j k l}, \as (\mu^2) \right) \, ,
\label{FullGamma}
\eeq
where $s_{ij} = 2 p_i \cdot p_j$, $\Gamma_n^{\rm dip}$ contains only two-particle 
correlations, and $\Delta_n$ is constructed out of quadrupole correlations, starting 
at three loops~\cite{Almelid:2015jia,Almelid:2017qju}, and constrained to depend 
on momenta only through the conformal-invariant cross ratios
\beq
  \rho_{i j k l} \, = \, \frac{p_i \cdot p_j \, p_k \cdot p_l}{p_i \cdot p_l \, p_j \cdot p_k} \, .
\label{cicrs}
\eeq
Up to two loops, only the dipole part of the infrared anomalous dimension matrix
is relevant. It can be written as~\cite{Gardi:2009qi,Gardi:2009zv,Becher:2009cu,
Becher:2009qa}
\beq
  \Gamma_n^{\rm dip} \left(\frac{s_{ij}}{\mu^2}, \as (\mu^2) \right) & = & 
  - \frac{1}{2} \, \widehat{\gamma}_K \left( \as (\mu^2) \right) \sum_{i=1}^n\sum_{j = i+1}^n
  \log \left( \frac{- s_{i j} - {\rm i} \varepsilon}{\mu^2} \right) {\bf T}_i \cdot {\bf T}_j 
  \nonumber \\ && \hspace{2cm} 
  + \, \sum_{i = 1}^n \gamma_i \left( \as(\mu^2) \right) \, ,
\label{GammaDip}
\eeq
where $\gamma_i$ is a collinear anomalous dimension, dependent on 
particle spin and related to the corresponding field anomalous dimension. 
The operators ${\bf T}_i$ act as `gluon insertion' operators, in a manner 
dependent on the colour representation of the hard particle $i$, as discussed 
in~\cite{Bassetto:1984ik,Catani:1996vz}. The coefficient of the logarithmic 
term is extracted from the light-like cusp anomalous dimension for colour 
representation ${\rm r}$, $\gamma_K^{\rm r} (\as)$, assuming that $\gamma_K^{\rm r} 
(\as) = C_{\rm r} \, \widehat{\gamma}_K ( \as )$, and dropping the quadratic Casimir 
eigenvalue $C_{\rm r}$: this assumption (`Casimir scaling') is known to be valid 
up to three loops, while there is solid numerical evidence that it breaks 
down at four loops, due to the presence of fourth-order Casimir
invariants~\cite{Moch:2017uml,Boels:2017ftb}.

Eqs.~(\ref{FullGamma}) and~(\ref{GammaDip}) highlight several remarkable
simplifications in the general structure of infrared poles: first of all, exponentiation
ties together different orders in perturbation theory in a non-trivial way; furthermore,
one observes that correlations involving three coloured particles are absent at 
NNLO at the level of the soft anomalous dimension, and can only arise in amplitudes 
through the mixing of one- and two-loop effects upon expanding the exponential; 
finally, to all orders in perturbation theory, non-dipole corrections are severely 
constrained to depend on momenta only through the variables in \eq{cicrs}. 
We expect these simplifying features to be reflected in the detailed structure 
of real radiation, and our goal is to set up tools to uncover and implement
these simplification. In order to proceed, we note that the compact expression 
in \eq{RGsol} is not sufficiently detailed to extract information relevant to the 
subtraction problem, where it is important to distinguish the contributions of 
soft and collinear configurations, and to understand the issue of double counting 
of soft-collinear poles. It is therefore necessary to take a step back to the full 
factorisation formula underlying \eq{RGsol}, which can be written as~\cite{Sen:1982bt,
Collins:1989bt,Sterman:1995fz,Catani:1998bh,Sterman:2002qn,Dixon:2008gr,
Gardi:2009qi,Gardi:2009zv,Becher:2009cu,Becher:2009qa,Feige:2014wja}
\beq
  {\cal A}_n \! \left( \frac{p_i}{\mu} \right) = \prod_{i = 1}^n  \left[
  \frac{{\cal J}_i \Big( (p_i \cdot n_i)^2/(n_i^2 \mu^2) 
  \Big)}{{\cal J}_{i, \, \E} \Big( (\beta_i \cdot n_i)^2/n_i^2 
  \Big)} \right] {\cal S}_n \left( \beta_i \cdot \beta_j \right) {\cal H}_n 
  \left( \frac{p_i \cdot p_j}{\mu^2}, \frac{(p_i \cdot n_i)^2}{n_i^2 \mu^2} \right) ,
\label{factoramp}
\eeq
where for simplicity we suppressed the dependence on the renormalised coupling
$\as (\mu^2)$ and on the regulator $\eps$. In \eq{factoramp}, the colour vector 
${\cal H}_n$ is a finite remainder (related, but not equal to ${\cal F}_n$ in 
\eq{IRfact}). For each hard massless particle with momentum $p_i$, we 
introduced a four-velocity vector $\beta_i$, $\beta_i^2 = 0$, obtained by rescaling 
$p_i$ by an arbitrary hard scale, say $\beta_i = p_i/\mu$, and a `factorisation 
vector' $n_i$, $n_i^2 \neq 0$, responsible for isolating the collinear region for 
particle $i$, and in order to enforce the gauge invariance of the collinear factors. 
For each hard particle, the {\it jet function} ${\cal J}_i$ collects all collinear 
singularities associated with the direction defined by $p_i$. The jet functions 
are spin dependent, and defined in terms of gauge-invariant matrix elements 
of fields and Wilson lines. For outgoing quarks with momentum $p$ and spin 
polarisation $s$ one defines
\beq
  \overline{u}_s (p) \, {\cal J}_q \! \left( \frac{(p \cdot n)^2}{n^2 \mu^2} \right) \, = \, 
  \langle p, s \, | \, \overline{\psi} (0) \, \Phi_n (0, \infty) \,  | 0 \rangle \, ,
\label{Jqdef}
\eeq
where the Wilson line operator is
\beq
  \Phi_v (\lambda_2, \lambda_1) \, \equiv \, {\cal P} \exp \left[ \, {\rm i} g_s 
  \int_{\lambda_1}^{\lambda_2} d \lambda \, v \cdot A (\lambda v) \right] \, .
\label{Wline}
\eeq
For (outgoing) gluons with momentum $k$ and polarisation $\lambda$, the definition
is more delicate, due to the requirement of gauge invariance: a straightforward
substitution of a gluon field for the quark field in \eq{Jqdef} is not satisfactory,
due to the non-homogeneous term in the gluon gauge transformation. The issue
has been well understood for a long time, initially in the context of giving
operator definitions of parton distribution functions for gluons~\cite{Collins:1989gx}.
In that case, the requirement is to find a gauge invariant quantity reducing 
to a gluon number operator in a physical gauge; a possible solution is to
use a particular projection of a field strength operator in place of the gluon
field in the equivalent of \eq{Jqdef}: the homogeneous gauge transformation
of the field strength can then be compensated by the Wilson line insertion.
At amplitude level, an elegant proposal was put forward in the context of SCET
in \cite{Becher:2009th,Becher:2010pd}, and we will use it in what follows.
We define
\beq 
  g_s \, \varepsilon^{* \, (\lambda)}_\mu (k) {\cal J}_{g}^{\mu \nu} 
  \! \left( \frac{(k \cdot n)^2}{n^2 \mu^2} \right) \, \equiv \, 
  \bra{k, \lambda} \, \Big[ \Phi_n (\infty, 0) \, {\rm i} D^\nu \, 
  \Phi_n(0, \infty) \, \Big] \ket{0} \, ,
\label{Jgdef}
\eeq
where we have not displayed colour indices, the covariant derivative $D_\mu = 
\partial_\mu - {\rm i} g_s A_\mu$ is evaluated at $x = 0$, and the extra power 
of $g_s$ on the left-hand side compensates for the effect of differentiating the 
Wilson line. 

We note that jet functions are single-particle quantities and do not carry 
any colour correlations from the full amplitude: the fact that collinear poles
have this property is a highly non-trivial consequence of gauge invariance
and diagrammatic power counting. Colour-correlated singularities arise only
from soft gluons, which, at leading power in their momentum, cannot transfer
energy between hard particles, but induce long-range colour mixing. The 
soft factor ${\cal S}_n$ is therefore a colour operator, defined in terms of 
semi-infinite light-like Wilson lines radiating out of the hard collision, each 
along the classical trajectory of one of the hard particles. One defines
\beq
  {\cal S}_n \left(\beta_i \cdot \beta_j \right) \, = \, \langle 0 |  \, \prod_{k = 1}^n 
  \Phi_{\beta_k} (\infty, 0) \, | 0 \rangle \, ,
\label{softcorr}
\eeq
where $\beta_i$ is the dimensionless four-velocity of the $i$-th hard particle, 
and where, for simplicity, we do not display the color indices of the Wilson lines.

Gluons that are both soft and collinear to one of the hard coloured particles
are present both in the jet functions, \eq{Jqdef} and \eq{Jgdef}, and in the 
soft matrix, \eq{softcorr}, and are therefore counted twice. It is however 
straightforward to subtract this double counting, since the soft approximation
of the jet function is simply given by the {\it eikonal jet}~\cite{Dixon:2008gr}
\beq
  {\cal J}_{\E} \left( \frac{(\beta \cdot n)^2}{n^2} \right)  \, = \, 
  \langle 0 | \, \Phi_{\beta}(\infty, 0) \, 
  \Phi_{n} (0, \infty) \, | 0 \rangle \, ,
\label{calJdef}
\eeq
and soft poles cancel in the ratio of the full jet to the eikonal jet, separately 
for each hard particle. This simple pattern of cancellation for soft-collinear
regions (which in particular does not contain any colour correlations) will 
be reflected in the structure of local counterterms for real radiation.

We conclude this section with two technical remarks. First, we note that 
the requirement that $n_i^2 \neq 0$ for all jet and eikonal jet functions is
designed in order to avoid the presence of spurious collinear divergences
associated with emissions from the $n_i$ Wilson lines. In practical calculations,
however, it is highly economical to take the $n_i^2 \to 0$ limit, provided
one can precisely control the contributions of spurious poles\footnote{For
a discussion of this point, see~\cite{Bonocore:2015esa}.}. Finally, we note 
that in dimensional regularisation all correlators of (semi-)infinite Wilson lines
are computed in perturbation theory in terms of scaleless integrals, 
which vanish in dimensional regularisation, so that the bare soft matrix
and eikonal jets equal unity. One can therefore extract the infrared poles 
of these matrix elements by computing their ultraviolet divergences, which 
allows to make use of standard renormalisation group arguments.
In practice, calculations can be performed with auxiliary regulators for
soft and collinear poles: one may for example tilt the $\beta_i$ Wilson 
lines off the light cone, and introduce a suppression for gluon emission at 
large distances, as done for example in~\cite{Gardi:2013saa,Falcioni:2014pka}.
General theorems then guarantee~\cite{Mitov:2010rp,Gardi:2011yz,Gardi:2013ita} 
that the resulting anomalous dimensions are independent of the chosen collinear 
and soft regulators.

%%%%%%%%%%%%%%%%%%%%%%%%%%%%%%%%%%%%%%%

\section{Subtraction procedures at NLO and NNLO}
\label{Subtra}

We now provide a brief description of a subtraction procedure at NLO and 
NNLO, for the case of massless coloured particles in the final state, identifying 
the local counterterms required in this case. Our goal here is to present 
the general structure of the procedure, which is sufficient for the purposes 
of the present paper: a detailed construction of a complete subtraction 
algorithm for this case is presented in~\cite{Magnea:2018hab}. 

Let us begin by establishing some notation. Given a scattering amplitude 
with $n$ massless particles in the final state, we write
\beq
  {\cal A}_n (p_i) \, = \, {\cal A}_n^{(0)} (p_i) \, + \, {\cal A}_n^{(1)} (p_i) \, + \,
  {\cal A}_n^{(2)} (p_i) \, + \, \ldots \, ,
\label{pertexpA}
\eeq
where ${\cal A}_n^{(0)} (p_i)$ is the Born amplitude for the process at hand 
(which may of course already contain powers of the strong coupling), while 
${\cal A}_n^{(k)} (p_i)$ is the $k$-loop correction. Given an infrared-safe 
observable $X$, one can then construct the perturbative expansion for 
the differential distribution of $X$, as
\beq
  \frac{d \sig}{d X} \, = \, \frac{d \sig_\LO}{d X} \, + \, 
  \frac{d \sig_\NLO}{d X} \, + \, 
  \frac{d \sig_\NNLO}{d X} \, + \, \ldots \, .
\label{pertexpsig}
\eeq
At each non-trivial order in perturbation theory, the differential distribution 
contains contributions with different numbers of final state particles, and 
the cancellation of infrared singularities takes place upon integration
over the phase space of unresolved radiation. Denoting with $d \Phi_m$ 
the Lorentz-invariant phase space measure for $m$ massless final state 
particles, and assuming that the observable involves $n$ particles at 
Born level, one can write in more detail
\beq
  \frac{d \sig_\LO}{dX} & = & \int d \Phi_n \, B_n \, \delta_n (X) \, , 
  \nonumber \\ [3pt]
  \frac{d \sig_\NLO}{d X} & = & \lim_{d \to 4} 
  \Bigg\{ \! \int d \Phi_n \, V_n \, \delta_n (X) + 
  \int d \Phi_\npo \, R_\npo \, \delta_\npo (X) \Bigg\} \, , 
  \label{pertO} \\ [3pt]
  \frac{d \sig_\NNLO}{d X} & = & \lim_{d \to 4} 
  \Bigg\{ \! \int d \pn \, VV_n \, \delta_n (X) + \int d \pnpo \, 
  RV_\npo \, \delta_\npo (X) \nonumber \\
  && \hspace{2cm} + \int d \pnpt \, RR_\npt \, \delta_\npt  (X) 
  \Bigg\} \, , \nonumber 
\eeq
where $\delta_m (X) \equiv \delta (X - X_m)$ fixes $X_m$, the expression 
for the observable appropriate for an $m$-particle configuration, to the 
prescribed value $X$. The integrands of the various terms can be 
expressed in terms of the squared scattering amplitudes involving 
$n$, $\npo$ and $\npt$ particles as
\beq
  && \hspace{2mm} B_n \, = \, \left| {\cal A}_n^{(0)} \right|^2 \, , \qquad 
  R_\npo \, = \, \left| {\cal A}_\npo^{(0)} \right|^2 \, , \qquad
  RR_\npt \, = \, \left| {\cal A}_\npt^{(0)} \right|^2 \, , \nonumber \\ [3pt]
  && V_n \, = \, 2 {\bf Re} \left[ {\cal A}_n^{(0) *} \, {\cal A}_n^{(1)} 
  \right] \, , \qquad
  VV_n \, = \, \left| {\cal A}_n^{(1)} \right|^2 \, + \, 2 {\bf Re} 
  \left[ {\cal A}_n^{(0) *} \, 
  {\cal A}_n^{(2)} \right] \, , \nonumber \\ [5pt]
  && \hspace{3.05cm} RV_\npo \, = \, 2 {\bf Re} 
  \left[ {\cal A}_\npo^{(0) *} \, {\cal A}_\npo^{(1)} \right] \, ,
\label{pertA2}
\eeq  
where unobserved quantum numbers (such as colour) not affecting the 
observable $X$  have been implicitly summed over. As briefly discussed 
in the Introduction, the problem of subtraction arises because the expressions 
$X_m$ for typical observables in the $m$-particle phase space, as well as the
corresponding matrix elements, are very 
intricate, requiring numerical integrations of the real emission contributions. 
It is then often necessary to perform the cancellation of infrared poles 
analytically, before turning to numerical tools. The subtraction approach 
proceeds by seeking approximations to the real-radiation matrix 
elements which must be accurate at leading power in the appropriate 
variables (for instance, energies or transverse momenta) in all singular regions. To 
be more precise, let us first consider the NLO distribution. In that case, 
we seek a {\it local counterterm} function $K_\npo$ in the $(\npo)$-particle 
phase space, with the requirement that it reproduces the singular 
behaviour of the real-radiation transition probability $R_\npo$ in all infrared 
limits, and, in our approach, with the further requirement that it should 
have a minimal degree of complexity, in order to allow for a direct 
analytic integration. Given such a function, we define the 
integrated NLO counterterm as
\beq
  I_n \, = \, \int d \Phi_{\rm rad}  \, K_\npo  \, ,
\label{intcountNLO}
\eeq
where we introduced the single-particle phase space measure $d 
\Phi_{\rm rad} = d \Phi_{\npo}/d \Phi_n$. We can now subtract the local 
counterterm $K_\npo$ from the real-emission probability $R_\npo$, 
obtaining an integrable function in the ($\npo$)-particle phase space, and 
then add back to the distribution the integrated counterterm $I_n$, 
which must cancel the explicit poles of the NLO virtual correction $V_n$. 
The result is
\vspace{3pt}
\beq
  \frac{d \sig_\NLO}{d X} & = & \int d \pn \Big( V_n + I_n \Big) \, 
  \delta_n (X) \nonumber \\ && \hspace{1cm} + \, 
  \int d \Phi_\npo \, \bigg[  R_\npo \, \delta_\npo (X) \, - \, K_\npo \,\,
  \delta_n (X) \bigg] \, .
\label{subtNLO}
\eeq
Note that no approximation has been introduced in passing from the second 
line of \eq{pertO} to \eq{subtNLO}. Thanks to the infrared safety of the observable 
$X$, the integrand in the second line of \eq{subtNLO} is now integrable 
everywhere in the $(\npo)$-particle phase space, and, at the same time, the
first line is free of infrared poles. The differential distribution in this form is 
therefore amenable to a direct numerical evaluation.

At NNLO, the cancellation pattern is considerably more intricate, but an exact 
subtraction procedure can still be constructed. At this order, infrared singularities
arise in three different configurations: in the double-radiation transition 
probability $RR_\npt$, either one or two emitted particles can become 
unresolved, and in the real-virtual transition probability $RV_\npo$ the 
single emitted particle  can similarly become unresolved. It is therefore
necessary to define three local counterterms: a function $K^{\two}_\npt$
in the $(\npt)$-particle phase space, approximating $RR_\npt$ in all 
singular regions with two unresolved particles, a function $K^{\one}_\npt$
in the $(\npt)$-particle phase space, approximating $RR_\npt$ in all 
singular regions with one unresolved particle, and a function $K^{\RV}_\npo$,
in the $(\npo)$-particle phase space, approximating $RV_\npo$ in all 
singular regions where the radiated particle becomes unresolved. It is
furthermore appropriate to separate the double-unresolved
counterterm as
\beq
  K^{\two}_\npt \, = \, K^{\otwo}_\npt + K^{\ttwo}_\npt \, ,
\label{splitdouble}
\eeq
where the first term collects all double-unresolved limits which are reached
hierarchically, with the first particle becoming unresolved at a faster rate than
the second one, while the second term contains all remaining double-unresolved
contributions, where the two particles become unresolved at the same rate (for
a detailed discussion of how to achieve this separation, see~\cite{Magnea:2018hab}).
One may then define the respective radiation phase spaces as
\beq
  d \Phi_{{\rm rad}, 1} \, = \, d \Phi_{\npt}/d \Phi_\npo \, , \quad
  d \Phi_{{\rm rad}, 2} \, = \, d \Phi_{\npt}/d \Phi_n \, , \quad 
  d \Phi_{\rm rad} \, = \, d \Phi_{\npo}/d \Phi_n \, , 
\label{radphsp}
\eeq
and introduce the integrated counterterms as  
\beq
  I^{\one}_\npo & = & \int d \Phi_{ {\rm rad}, \, 1} \, K^{\one}_\npt \, , 
  \qquad \hspace{7pt}  I^{\otwo}_\npo \, = \, 
  \int d \Phi_{ {\rm rad}, \, 1} \, K^{\otwo}_\npt \, ,
  \nonumber \\ [3pt]
  I^{\ttwo}_n & = & \int d \Phi_{ {\rm rad}, \, 2} \, K^{\ttwo}_\npt \, ,  
  \qquad I^{\RV}_n \, = \, \int d \Phi_{\rm rad} \, K^{\RV}_\npo \, .
\label{intcountNNLO}
\eeq
As was the case at NLO, in \eq{intcountNLO}, also in \eq{intcountNNLO} the 
subscripts indicate the number of particles whose phase space still needs 
to be integrated. Specifically, $I^{\ttwo}_n$ and $I^{\RV}_n$ depend on 
the Born phase-space configuration, with all $n$ particles resolved, and 
contain explicit infrared poles that cancel those of the double virtual 
transition probability $VV_n$. On the other hand, $I^{\one}_\npo$ 
depends on the phase space variables of $(\npo)$ particles, and has 
explicit infrared poles cancelling those of the real-virtual transition 
probability $RV_\npo$; the resulting finite combination, however, 
can still have singular limits when the radiated particle becomes unresolved:
those singular limits must be subtracted by combining $K^{\RV}_\npo$
with $I^{\otwo}_\npo$, in order to cancel the respective explicit poles.
Our final expression for the subtracted NNLO distribution is therefore
\beq
  \frac{d \sig_\NNLO}{dX} & = & \int d \Phi_n
  \Big[ VV_n + I^{\ttwo}_n + I^{\RV}_n \Big] \, \delta_n (X) 
  \label{subtNNLO} \\
  & + & \int d \Phi_\npo \bigg[ \left( RV_{\npo} + I^{\one}_{\npo} \right)
  \delta_\npo (X) -  \left( K^{\RV}_\npo - I^{\otwo}_\npo \right)
  \delta_n (X) \bigg] \nonumber \\
  & + & \int d\Phi_\npt \bigg[ RR_\npt \, \delta_\npt (X) - K^{\one}_\npt \,\,
  \delta_\npo (X) - \left( K^{\otwo}_\npt + K^{\ttwo}_\npt \right) \delta_n (X) 
  \bigg] \, . \nonumber
\eeq
One verifies that no approximation has been made in going from 
the third line of \eq{pertO} to \eq{subtNNLO}. Furthermore, each line in 
\eq{subtNNLO} is both finite in four dimensions, and integrable in the 
respective phase spaces.

Clearly, \eq{subtNNLO} is only the starting point in the construction of 
a full-fledged subtraction algorithm: the next crucial step is the explicit
definition of the necessary local counterterms, which must properly
organise all soft, collinear and soft-collinear regions avoiding double 
counting; in the process, it is necessary to construct precise phase-space 
mappings in order to exactly factorise radiative from non-radiative phase spaces;
finally, the local counterterms must be analytically integrated
in the respective radiation phase spaces. In the remainder of this paper, 
we discuss a systematic construction of the local counterterms, which 
we will carry out explicitly up to NNLO, but which is applicable in principle 
at any perturbative order. A detailed algorithmic implementation of 
\eq{subtNNLO} for final-state massless partons has been presented 
in~\cite{Magnea:2018hab}. In what follows, our main concern is not the 
calculation of NNLO kernels, which have been known for a long 
time~\cite{Catani:1999ss,Campbell:1997hg,Bern:1999ry,Catani:2000pi}: 
rather, we plan to show how information from the factorisation of virtual 
corrections allows to organise and simplify the NNLO subtraction 
procedure, pointing to possible future extensions to higher perturbative 
orders.

%%%%%%%%%%%%%%%%%%%%%%%%%%%%%%%%%%%%%%%

\section{Local counterterms for soft real radiation}
\label{Softct}

Our general strategy to define local counterterms is to construct eikonal 
form factors and radiative jet functions including real radiation: these 
functions, when integrated over the final-state phase space and combined
with their virtual counterparts using completeness relations, build up eikonal 
and collinear total cross sections, which are finite by the general theorems 
of Refs.~\cite{Bloch:1937pw,Kinoshita:1962ur,Lee:1964is,Grammer:1973db}.
Let us begin with the case of purely soft final state radiation (which of 
course includes soft-collinear particles as well). Considering $n$ hard
particles, represented by Wilson lines in the soft approximation, radiating
$m$ soft gluons, we define the {\it eikonal form factor}
\beq
  {\cal S}_{n, \, m} \left(k_1, \ldots, k_m; \beta_i \right)
  & \equiv & \bra{k_1, \lambda_1; \ldots; k_m, \lambda_m} \, 
  \prod_{i = 1}^n \Phi_{\beta_i} (\infty, 0) \, \ket{0} \nonumber \\
  & \equiv & \epsilon_{\mu_1}^{* \, (\lambda_1)} (k_1) \ldots 
  \epsilon_{\mu_m}^{* \, (\lambda_m)} (k_m) \, 
  J_{\cal S}^{\mu_1 \ldots \mu_m}
  \left(k_1, \ldots, k_m; \beta_i \right) \nonumber \\
  & \equiv &  \sum_{p = 0}^\infty \,
  {\cal S}_{n, \, m}^{(p)} \left(k_1, \ldots, k_m; \beta_i \right) \, ,
\label{softrad}
\eeq
where in the second line we have defined multiple soft gluon currents
$J_{\cal S}^{\mu_1 \ldots \mu_m}$, in the third line we have introduced
the perturbative expansion of the form factors, and we are not displaying 
colour indices to simplify the notation. A well known property of the soft 
approximation at leading power in the soft momenta is spin-independence: 
thus the multiple soft gluon currents are independent of the gluon 
polarisations $\lambda_i$, and the definition easily generalises to the 
emission of final state soft fermions. Note that at this stage the form 
factor contains loop corrections to all orders in perturbation theory.

Our underlying assumption is that the exact amplitude for the emission
of $m$ soft gluons (which may in turn radiate soft quark-antiquark pairs)
from $n$ hard coloured particles obeys, to all orders, the factorisation
\beq
  {\cal A}_{n, \, m} \left(k_1, \ldots, k_m; p_i \right)
  \, =  \, {\cal S}_{n, \, m} \left(k_1, \ldots, k_m; 
  \beta_i \right) \, {\cal H}_n (p_i) \, + \, {\cal R}_{n, \, m} \, ,
\label{ampfact}
\eeq
where the remainder ${\cal R}_{n, \, m}$ is finite in four dimensions, and
integrable in the soft particle phase space. After renormalisation, the
amplitude ${\cal A}_{n, \, m}$ is ultraviolet finite, and all virtual soft
poles, as well as all contributions that are non-integrable in the soft
particle phase space, are contained in the soft form factor ${\cal S}_{n, 
\, m}$. \eq{ampfact} is proven to all orders for $m = 0$, and it is consistent 
with all known perturbative results, in particular with the arguments 
of~\cite{Catani:1999ss,Campbell:1997hg,Catani:2000pi}; a formal 
all-order proof has however not yet been provided: we treat it as a 
working assumption, which is known to be correct at NNLO.

Squaring \eq{ampfact}, and performing the trivial helicity sum, one finds,
at leading-power in the soft momenta
\beq
  \sum_{\{\lambda_i\}} \left| {\cal A}_{n, \, m} \left(k_1, \ldots, k_m; p_i 
  \right) \right|^2 \,\, \simeq \,\, {\cal H}_n^\dagger (p_i) \, 
  S_{n, \, m} \left(k_1, \ldots, k_m; \beta_i \right) \, {\cal H}_n (p_i) \, ,
\label{squampfact}
\eeq
where we introduced the {\it eikonal transition probability}
\beq
  && S_{n, \, m} \left(k_1, \ldots, k_m; \beta_i \right) \, \equiv \, 
  \sum_{p = 0}^\infty \, S_{n, \, m}^{(p)} \left(k_1, \ldots, k_m; \beta_i \right) 
  \label{softsigma} \\
  && \equiv \,\, 
  \sum_{\{\lambda_i\}} 
  \bra{0} \prod_{i = 1}^n 
  \Phi_{\beta_i} (0, \infty) \ket{k_1, \lambda_1; \ldots; k_m, \lambda_m}
  \bra{k_1, \lambda_1; \ldots; k_m, \lambda_m} \prod_{i = 1}^n 
  \Phi_{\beta_i} (\infty, 0) \ket{0} \, , \nonumber
\eeq
for fixed final-state soft momenta $k_i$. \eq{softsigma} provides a natural 
definition of local soft counterterms, order by order in perturbation theory: 
indeed integrating over the soft particle phase space for fixed $m$, and then
summing over $m$, one can use completeness to get
\beq
  \sum_{m = 0}^\infty \int d \Phi_m \, S_{n, \,m} \left(k_1, \ldots, k_m; 
  \beta_i \right) \, = \, 
  \bra{0} \prod_{i = 1}^n 
  \Phi_{\beta_i} (0, \infty) \prod_{i = 1}^n 
  \Phi_{\beta_i} (\infty, 0) \ket{0} \, .
\label{complete}
\eeq
\eq{complete}, up to simple modifications\footnote{For example, if the 
$m$-particle phase space includes a momentum-conservation $\delta$-function 
setting the total final state energy to a fixed value $\mu$, which is irrelevant 
in the present context, the constraint can be implemented by shifting the origin 
of one of the two sets of Wilson lines on the {\it r.h.s.} of \eq{complete} in a 
timelike direction by an amount $\lambda$, and introducing a Fourier transform 
with a weight $\lambda \mu$. Notice that operator products in all our matrix 
elements are understood to be time ordered when needed.}, can be interpreted 
as an eikonal total cross section. When all coloured particles are in the final 
state, such a cross section is finite to all orders by the standard cancellation 
theorems (which can be verified by explicit power counting); with initial state 
colour, the eikonal cross section is affected by collinear divergences which 
can be treated by conventional collinear factorisation~\cite{Laenen:2000ij}: 
indeed, in our framework, these collinear divergences are included in eikonal 
jet factors to be discussed in \secn{Collct}. As far as soft divergences are 
concerned, we conclude that the kernels $S_{n, \,m}$ provide completely local 
soft approximations to the relevant squared matrix element, valid at leading 
power in the soft momenta, and they cancel the virtual soft poles order by 
order in perturbation theory: this identifies them as candidate counterterms 
for subtraction in the soft sector.

Let us now illustrate this general framework with simple examples, 
recovering known results at low orders. A classic case in point is 
single-gluon emission from a multi-particle configuration at tree level. 
\eq{ampfact} for $m = 1$ and at lowest order reads
\beq
  {\cal A}^{(0)}_{n, \, 1} (k, p_i) \, = \, \epsilon^{* \, (\lambda)} (k) \cdot 
  J_{\cal S}^{(0)} (k, \beta_i) \, {\cal H}^{(0)}_n (p_i) + {\cal O}(k^0)  \, ,
\label{treesoft}
\eeq
with the definition
\beq
  \epsilon^{* \, (\lambda)} (k) \cdot J_{\cal S}^{(0)} (k, \beta_i) \, = \, 
  {\cal S}_{n, \, 1}^{(0)} \left(k; \beta_i \right) \, = \,  
  \left. \bra{k, \lambda} \, \prod_{i = 1}^n \Phi_{\beta_i} (\infty, 0) \, 
  \ket{0} \right|_{\rm tree}  \, .
\label{treesoftcurr}
\eeq
Explicit calculation expanding the Wilson-line operators in powers 
of the coupling, or directly with eikonal Feynman rules, easily yields 
the well-known result for the tree-level soft-gluon emission 
current~\cite{Bassetto:1984ik,Catani:1999ss}
\beq 
  J_{\cal S}^{\mu \, (0)} (k; \beta_i) \, = \, g_s \, \sum_{i = 1}^n \,
  \frac{\beta_i^\mu}{\beta_i \cdot k} \, {\bf T}_i  \, \, .
\label{treesoftcurexp}
\eeq
Squaring the tree-level amplitude one finds the leading-power transition 
probability
\beq
  \sum_\lambda \left| {\cal A}^{(0)}_{n, \, 1} (k, p_i) \right|^2 
  & \simeq & {\cal H}^{(0) \, \dagger}_n (p_i)  \, S_{n, \, 1}^{(0)} 
  \left(k; \beta_i \right) \, {\cal H}^{(0)}_n (p_i) \nonumber \\ 
  & = & - \, 4 \pi \alpha_s \, \sum_{i,j = 1}^n \, 
  \frac{\beta_i \cdot \beta_j}{\beta_i \cdot k \, \beta_j \cdot k} \,\,  
  {\cal A}^{(0) \dagger}_n (p_i) \, {\bf T}_i \cdot {\bf T}_j \, 
  {\cal A}^{(0)}_n (p_i)  \, ,
\label{treesqu}
\eeq
where we used the fact that at tree level there is no need to distinguish 
between ${\cal H}^{(0)}_n$ and ${\cal A}^{(0)}_n$; we recognise the 
colour-correlated Born probability, multiplied times the standard eikonal 
prefactor. Multiple soft-particle radiation at tree level is similarly easy 
to compute: for the case of two gluons, one directly recovers the result 
of~\cite{Catani:1999ss}
\beq 
  \left[ J_{\cal S}^{(0)} \right]_{\mu_1 \mu_2}^{a_1 a_2} 
  (k_1, k_2; \beta_i) & = & 4 \pi \as \, \Bigg\{ \sum_{i = 1}^n \, 
  \left[ \beta_{i, \, \mu_1} \beta_{i, \, \mu_2} \left(
  \frac{T_i^{a_2} T_i^{a_1}}{\beta_i \cdot k_2 \, \beta_i \cdot 
  \left( k_1 + k_2 \right)} \, + \, \left( 1 \leftrightarrow 2 \right) \right) \right. 
  \nonumber \\
  && \hspace{-2cm} \left. - \, {\rm i} f^{\, \, \, a_1 a_2}_a \, T_i^a \,\,
  \frac{\beta_i \cdot \left( k_2 - k_1 \right) 
  g_{\mu_1 \mu_2} + 2 \beta_{i, \, \mu_1} k_{1, \, \mu_2} - 
  2 \beta_{i, \, \mu_2} k_{1, \, \mu_1}}{2 k_1 \cdot k_2 \, 
  \beta_i \cdot \left( k_1 + k_2 \right)} \right] \nonumber \\
  && \hspace{-2cm} + \, \sum_{i = 1}^n \sum_{j \neq i} 
  T_i^{a_1} \, T_j^{a_2} \, \frac{\beta_{i, \, \mu_1}}{\beta_i \cdot k_1} \, 
  \frac{\beta_{j, \, \mu_2}}{\beta_j \cdot k_2} \Bigg\} \, ,
\label{treesoftcurfull}
\eeq
with the last line representing uncorrelated emission from two different 
hard particles, and the first two lines collecting terms arising from 
double emission from a single hard particle. Currents corresponding 
to the radiation of soft quark-antiquark pairs, or for emissions with 
higher multiplicity, can similarly be computed directly in Feynman 
gauge in a straightforward manner.

At loop level, the organisation of counterterms becomes more 
interesting. Let us for example consider single-gluon emission at 
one loop: expanding \eq{ampfact} for $m = 1$ to first 
non-trivial order we find
\beq
  {\cal A}_{n, \, 1}^{(1)} \left(k ; p_i \right) \, = \, {\cal S}_{n, \, 1}^{(0)} 
  \left(k; \beta_i \right) \, {\cal H}^{(1)}_n (p_i) \, + \,  
  {\cal S}_{n, \, 1}^{(1)} \left(k; \beta_i \right) \, {\cal H}^{(0)}_n (p_i) \, .
\label{expone1}
\eeq
The first term corresponds to a tree-level soft-gluon emission multiplying 
the finite part of the one-loop correction to the Born process; in the 
second term the soft function is evaluated at one-loop, and therefore 
has both explicit soft poles and singular factors from single soft real 
radiation: it multiplies the Born amplitude. In this case, the proposed 
factorisation appears to differ from the one proposed in~\cite{Catani:2000pi}, 
which reads
\beq
  {\cal A}_{n, \, 1} \left(k ; p_i \right) \, \simeq \, 
  \epsilon^{* \, (\lambda)} (k) \cdot 
  J_{\CG} \left( k, \beta_i \right) \, {\cal A}_n (p_i) \, .
\label{twofact}
\eeq
Here the Catani-Grazzini soft current $J_{\CG} \! \left( k, \beta_i \right)$ 
multiplies the full $n$-particle amplitude, including loop corrections 
containing infrared poles, whereas in \eq{ampfact} for $m = 1$ the 
hard function ${\cal H}_n (p_i)$ is finite. It is, however, easy to map 
the two calculations, using \eq{ampfact} for $m = 0$, and solving for 
the one-loop hard part ${\cal H}_n^{(1)} (p_i)$. One finds
\beq
  {\cal H}^{(1)}_n (p_i) \, = \, {\cal A}^{(1)}_n \left( p_i \right) -
  {\cal S}^{(1)}_n \left( \beta_i \right) \, {\cal A}^{(0)}_n (p_i)  \, ,
\label{exptwo}
\eeq
where we normalised ${\cal S}^{(0)}_n$ to the identity operator in colour space.
This leads to an expression for the Catani-Grazzini one-loop soft-gluon
current in terms of eikonal form factors, as
\beq
  \epsilon^{* \, (\lambda)} (k)  \cdot J^{(1)}_{\CG} (k, \beta_i) \, 
  \,\, = \,\, {\cal S}_{n, \, 1}^{(1)} \left(k; \beta_i \right) \, - \,
  {\cal S}_{n, \, 1}^{(0)} \left(k; \beta_i \right) \, {\cal S}^{(1)}_n 
  \left( \beta_i \right) \, .
\label{CGolsc}
\eeq
Comparing \eq{CGolsc} with the calculation in~\cite{Catani:2000pi}, one 
easily recognises that the same combination of Feynman diagrams is 
involved, and one recovers the known result
\beq
  \left[ J^{(1)}_{\CG} \right]^\mu_a (k, \beta_i) & = & - \frac{\as}{4 \pi} \, 
  \frac{1}{\eps^2} \, \frac{\Gamma^3 (1 - \eps) 
  \Gamma^2 (1 + \eps)}{\Gamma(1 - 2 \eps)} \nonumber \\
  && \hspace{-2cm}  \times \, {\rm i} f_a^{\,\, b c} 
  \sum_{i = 1}^n \sum_{j \neq i} T_i^b T_j^c
  \left( \frac{\beta_i^\mu}{\beta_i \cdot k} - 
  \frac{\beta_j^\mu}{\beta_j \cdot k} \right) 
  \left[ \frac{2 \pi \mu^2 \left( - \beta_i \cdot \beta_j \right)}{\beta_i \cdot k 
  \, \beta_j \cdot k} \right]^\eps \, .
\label{olsc}
\eeq
Phrasing the calculation in terms of eikonal form factors allows for a 
straightforward and systematic generalisation to higher orders. For 
example, expanding \eq{ampfact}, for $m = 1$, to two loops, one finds
\beq
  {\cal A}_{n, \, 1}^{(2)} \left(k ; p_i \right)  & \simeq &  
  {\cal S}_{n, \, 1}^{(0)} \left(k; \beta_i \right) \, 
  {\cal H}^{(2)}_n (p_i) \, + \,  {\cal S}_{n, \, 1}^{(1)} \left(k; \beta_i \right) \, 
  {\cal H}^{(1)}_n (p_i)
  \nonumber \\
  && \hspace{2cm} + \,  {\cal S}_{n, \, 1}^{(2)} \left(k; \beta_i \right) \, 
  {\cal H}^{(0)}_n (p_i) \, .
\label{expone}
\eeq
The expression for ${\cal H}^{(1)}_n$ is given in \eq{exptwo}; furthermore,
one can similarly derive an expression for ${\cal H}^{(2)}_n$ from the 
two-loop expansion of \eq{ampfact} for $m = 0$, obtaining
\beq
  {\cal H}^{(2)}_n (p_i) & = & {\cal A}^{(2)}_n \left( p_i \right) -
  {\cal S}^{(1)}_n \left( \beta_i \right) \, {\cal A}^{(1)}_n (p_i) + 
  \left[ {\cal S}^{(1)}_n \left( \beta_i \right) \right]^2 \, {\cal A}^{(0)}_n (p_i) 
  \nonumber \\ && \hspace{5mm}
  \, - \,\,  {\cal S}^{(2)}_n \left( \beta_i \right) \, {\cal A}^{(0)}_n (p_i) \, .
\label{expthree}
\eeq
Substituting the expressions for the hard parts into \eq{expone}, and 
comparing with \eq{twofact}, one finds the two-loop soft-gluon current
\beq
  \epsilon^{* \, (\lambda)} (k)  \cdot J^{(2)}_{\CG} (k, \beta_i) \, 
  & = & {\cal S}_{n, \, 1}^{(2)} \left(k; \beta_i \right) - 
  {\cal S}_{n, \, 1}^{(1)} \left(k; \beta_i \right) {\cal S}^{(1)}_n 
  \left( \beta_i \right) 
  \nonumber \\ &&  \hspace{5mm}  
  - \, \, {\cal S}_{n, \, 1}^{(0)} \left(k; \beta_i \right) \left[ {\cal S}^{(2)}_n 
  \left( \beta_i \right) - \left( {\cal S}^{(1)}_n 
  \left( \beta_i \right) \right)^2 \right] \, .  
\label{CGosc2}
\eeq
Note that in expressions such as \eq{CGosc2} the ordering of factors 
is important, since the form factors ${\cal S}$ are colour operators. 
Note also that all terms in \eq{CGosc2}, except the first one, are already 
known for general massless $n$-point Born processes. The two-loop 
soft-gluon current was computed for $n = 2$ by extracting it from 
known two-loop matrix elements in Refs.~\cite{Badger:2004uk,Li:2013lsa,
Duhr:2013msa}. \eq{CGosc2} provides a precise framework 
for the calculation for generic processes with $n$ coloured particles 
at Born level. Clearly, it is not difficult to derive expression similar to 
\eq{CGosc2} for the case of multiple soft-gluon radiation at the desired
loop level.

%%%%%%%%%%%%%%%%%%%%%%%%%%%%%%%%%%%%%%%

\section{Local counterterms for collinear real radiation}
\label{Collct}

The strategy to define local collinear counterterms is very similar to the one
adopted in the soft case. We begin by
allowing for further final-state radiation in the operator matrix elements defining
the jet functions in \eq{Jqdef} and \eq{Jgdef}. This leads to the definition of
{\it radiative jet functions}, which are universal, but distinguish whether the 
emitting hard parton is a quark or a gluon. In particular, let us consider first 
a final state with a hard quark carrying momentum $p$ and spin $s$, and 
radiating $m$ gluons. In this case we define
\beq
  \overline{u}_s (p) \, {\cal J}_{q, \, m} \left(k_1, \ldots, k_m; p, n \right)
  & \equiv & \bra{p, s; k_1, \lambda_1; \ldots; k_m, \lambda_m}  \overline{\psi} (0) \, 
  \Phi_{n} (0, \infty) \ket{0} \nonumber
  \nonumber \\ 
  & \equiv & \overline{u}_s (p) \sum_{p = 0}^\infty 
  {\cal J}_{q, \, m}^{(p)} \left(k_1, \ldots, k_m; p, n \right) \, ,
\label{qradjet}
\eeq
where we extracted the quark wave function, so that ${\cal J}_{q, \, 0}$ coincides 
with the virtual quark jet defined in \eq{Jqdef}, and is normalised to unity at tree 
level. Gluon polarisation vectors, on the other hand, are still included in the function
${\cal J}_{q, \, m}$, and could be extracted to define collinear currents in a manner
analogous to what was done in \eq{softrad} for soft currents. The radiative quark jet 
function is gauge invariant in the same way as the non-radiative one discussed 
in \secn{Factor}: it is a matrix element involving only physical states, where the
gauge transformation properties of the field operator are compensated by the 
Wilson line; furthermore, like its non-radiative counterpart, it does not involve 
colour correlations with the other hard partons in process. The definition is 
valid to all orders in perturbation theory, and the second line of \eq{qradjet} gives 
the perturbative expansion, with ${\cal J}_{q, \, m}^{(p)}$ proportional to $g_s^{\, 2 p 
+ m}$. Notice however that the gluon momenta in \eq{qradjet} are unconstrained,
and collinear limits must be explicitly taken at a later stage in the calculation. 

At cross-section level, the definition of radiative jet functions is slightly more elaborate
than was the case for soft functions, since one must allow for non-trivial momentum
flow. This can be done in a standard way by shifting the position of the quark field
in the complex 
conjugate amplitude, and then taking a Fourier transform in order to fix the total 
momentum flowing into the final state, setting $l^\mu = p_i^\mu + \sum_{i=1}^m 
k_i^\mu $. In the unpolarised case, one may sum over polarisations and define
the cross-section-level radiative quark jet function as
\beq
\label{qradjetsq}
  && J_{q, \, m} \left(k_1, \ldots , k_m; l, p, n \right) \, \equiv \,
  \sum_{p = 0}^\infty J_{q, \, m}^{(p)} \left(k_1, \ldots , k_m; l, p, n \right) \\
  && \equiv \,
  \int d^d x \, {\rm e}^{{\rm i} l \cdot x} \, \sum_{\{ \lambda_j \}} \bra{0} 
  \Phi_n (\infty, x) \, \psi(x) \ket{p, s; k_j, \lambda_j}
  \bra{p, s; k_j, \lambda_j} \overline{\psi} (0) \, 
  \Phi_n (0, \infty) \ket{0} \, . \nonumber
\eeq
The perturbative coefficients $J_{q, \, m}^{(p)}$ of the radiative jet function 
$J_{q, \, m}$, computed in the collinear limit, provide natural candidates for
collinear counterterms, to any order in perturbation theory, as will be illustrated
below, in \secn{NLOres} at NLO and in \secn{NNLOres} at NNLO. 

For gluon-induced processes, we can proceed in the same way, starting with 
\eq{Jgdef}, and introducing the (amplitude-level) radiative gluon jet functions
as
\beq 
\label{collinear_current_glue}
  &&  g_s \, \varepsilon^{* \, (\lambda)}_\mu (k) {\cal J}_{g, \, m}^{\mu \nu} 
  \left( k_1, \ldots, k_m; k, n \right) \, \equiv \,
  g_s \, \varepsilon^{* \, (\lambda)}_\mu (k) \sum_{p = 0}^\infty 
  {\cal J}_{g, \, m}^{(p), \, \mu \nu} \left(k_1, \ldots, k_m; k, n \right) \\
  && \hspace{5cm} \equiv
  \bra{k, \lambda; k_1, \lambda_1; \ldots; k_m, \lambda_m}  
  \Phi_n (\infty, 0) \, {\rm i} D^\nu \, \Phi_n (0, \infty) \ket{0}  \, , \nn
\eeq
where again we are not displaying colour indices, and polarisation vectors
for the radiated gluons are included in the definition of ${\cal J}_{g, \, m}^{\mu 
\nu}$. The definition (\ref{collinear_current_glue}) can be used to construct a 
cross-section-level radiative gluon jet function, as was done for the quark. It reads
\beq 
\label{collinear_current_cross_level}
  && g_s^2 \, J_{g, \, m}^{\, \mu \nu} \left(k_1, \ldots, k_m; k, n \right)
  \equiv  g_s^2 \, \sum_{p = 0}^\infty J_{g, \, m}^{(p) , \, \mu \nu} \!
  \left( k_1, \ldots k_m; l, k, n \right) \\  \nn
  && \equiv \int d^dx \; {\rm e}^{{\rm i} l \cdot x } \, 
  \sum_{\{ \lambda_j \}} \bra{0} \left[ \Phi_n (\infty, x) \;  
  {\rm i} D^\mu \, \Phi_n(x, \infty) \right]^\dagger 
  \ket{k, \lambda; k_j, \lambda_j}  \\
  && \nn  \hspace{50mm} \bra{k, \lambda; k_j, \lambda_j} 
  \Phi_n (\infty, 0) \; {\rm i} D^\nu \, \Phi_n(0, \infty) \; \ket{0} \; .
\eeq
To illustrate the usefulness of radiative jet functions as collinear counterterms, 
let us focus, as an example, on the quark-induced jet function. In analogy to 
what was done in the soft sector, we note that summing over the number of 
radiated particles, and integrating over their phase space, by completeness 
one finds
\beq
  && \sum_{m = 0}^\infty \int d \Phi_{m+1} \, 
  J_{q, \, m} \left(k_1, \ldots, k_m; l, p, n \right) \nonumber \\ 
  && \hspace{2cm} = \, 
  {\rm Disc} \left[ \int d^d x \, {\rm e}^{{\rm i} l \cdot x} \,
  \bra{0} \Phi_n (\infty, x) \psi(x) \overline{\psi} (0) \Phi_n (0, \infty) \ket{0} \right] \, .
\label{compcoll}
\eeq
The {\it r.h.s.} of \eq{compcoll} gives the imaginary part of a generalised two-point 
function, which is a finite quantity, since it is fully inclusive in the final state. The 
$m = 0$ contribution contains the virtual collinear poles associated with an 
outgoing quark of momentum $p$, and therefore the real radiation contributions 
for $m \neq 0$, given by \eq{qradjetsq}, must cancel those poles order by order 
in perturbation theory, as desired. Inclusive cross-section-level jet functions 
such as the integrated quantity in \eq{compcoll} have been used in the context
of threshold resummations for many years, starting with the seminal papers
in Ref.~\cite{Collins:1981uk,Sterman:1987aj}.
\begin{figure}[t]
 \centering
\subfigure[]
 {\includegraphics[scale=0.4]{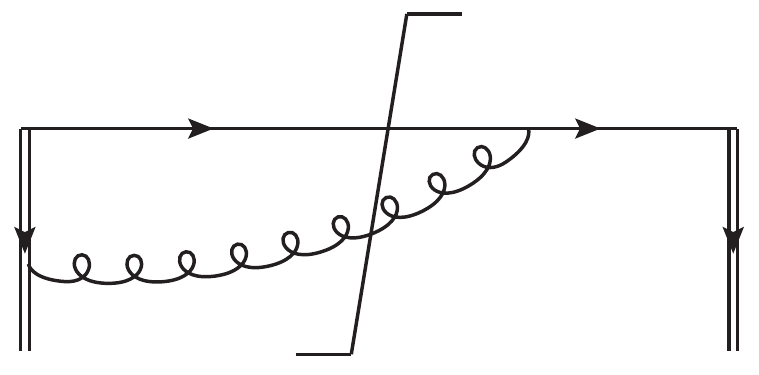}}
   \hspace{1cm}
\subfigure[]
 {\includegraphics[scale=0.4]{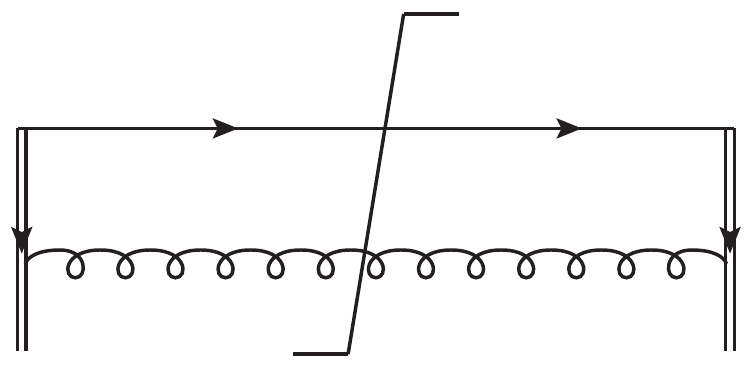}}
    \hspace{1cm}
\subfigure[]
 {\includegraphics[scale=0.4]{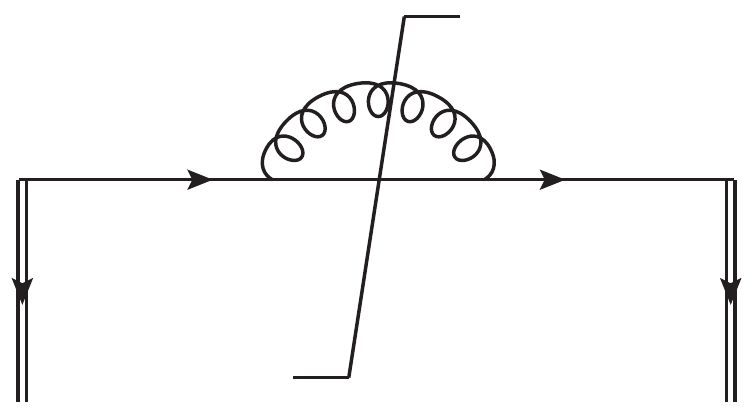}}
 \caption{One-loop contributions to cross-section-level radiative quark jet function}
 \label{fig:cut_one_loop_jet}
 \end{figure}

We can perform a simple test of the correctness of our method by computing the 
single-gluon radiative jet for an outgoing quark with momentum $p^\mu$. In 
Feynman gauge, the lowest perturbative order in the coupling constant receives 
contributions from three different diagrams, shown in Fig.~\ref{fig:cut_one_loop_jet}, 
which give the result
\beq
  \sum_s J_{q, \, 1} \left( k; l, p, n \right) & = & 
  \frac{4 \pi \alpha_s C_F}{(l^2)^2} \, (2 \pi)^d \delta^d \left(l - p - k \right) \nn \\
  && \hspace{3mm}  \times
  \left[ - \slash{l} \gamma_\mu \slash{p} \gamma^\mu \slash{l} 
  + \frac{l^2}{k \cdot  n}  \left( \slash{l} \slash{n} \slash{p} 
  + \slash{p} \slash{n} \slash{l} \right) \right] \, , 
\label{onecollglu}
\eeq
where $p^2 = k^2 = 0$, and up to corrections proportional to $n^2$.
It is easy to trace the contributions of the
three diagrams in Fig.~\ref{fig:cut_one_loop_jet} in the axial gauge calculation of 
Ref.~\cite{Catani:1999ss}. Notice however that in \eq{onecollglu} the collinear limit 
for $k$, corresponding to $l^2 \to 0$, has not been taken yet. This is easily achieved
by introducing a Sudakov parametrisation for momenta $p^\mu$ and $k^\mu$, and
taking the $k_\perp \to 0$ limit, setting
\beq
  p^\mu \, = \, z l^\mu + {\cal O} \left( l_\perp \right) \, , \qquad
  k^\mu  \, = \, (1 - z) l^\mu + {\cal O} \left( l_\perp \right) \, , \qquad
  n^2 \, = \, 0 \, .
\label{Sudakov}
\eeq
Due to the prefactor of order  $\mathcal{O}\left[(l_\bot^2 \right)^{-1}]$, the leading 
behaviour in the $l_\bot \rightarrow 0$ limit is recovered by setting $l_\bot = 0$ in 
the square bracket. This yields
\beq
  \sum_s J_{q, \, 1} \left( k; l, p, n \right) \, = \, 
  \frac{8 \pi \alpha_s C_F}{l^2} \, (2 \pi)^d \, \delta^d \left(l - p - k \right)
  \left[ \frac{1 + z^2}{1 - z} - \epsilon \left( 1 - z \right) \right] \, ,
\label{AP0}
\eeq
up to corrections of order $l_\bot$. In the square bracket, as expected, we recognise 
the leading order unpolarised DGLAP splitting function $P_{q \rightarrow qg}$.
\begin{figure}[t]
 \centering
\subfigure[]
 {\includegraphics[scale=0.4]{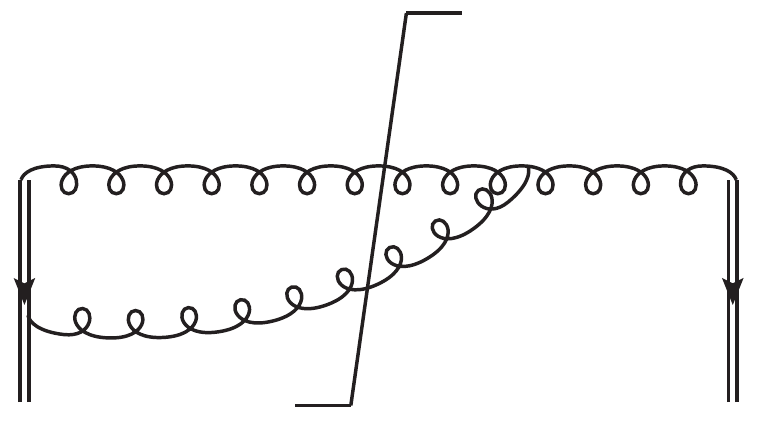}}
   \hspace{1cm}
\subfigure[]
 {\includegraphics[scale=0.4]{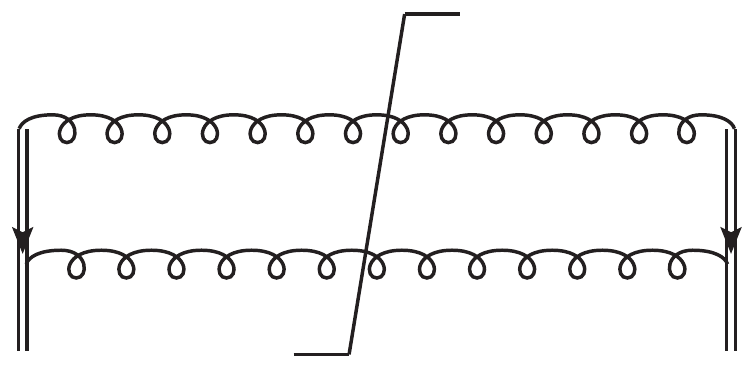}}
    \hspace{1cm}
\subfigure[]
 {\includegraphics[scale=0.4]{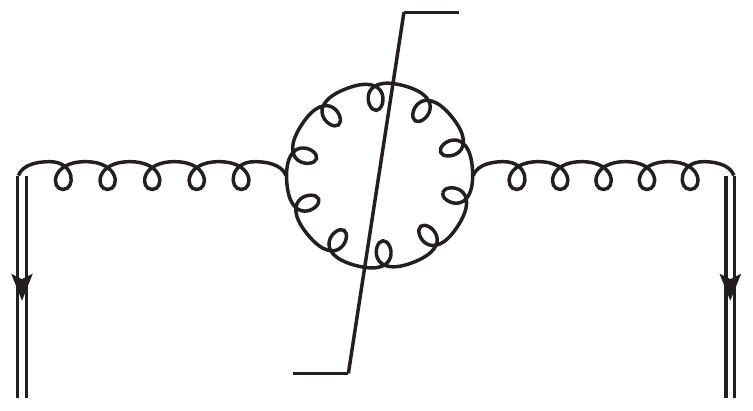}}
 \caption{One loop contributions to cross-section-level radiative gluon jet function}
 \label{fig:cut_one_loop_jet_glue}
 \end{figure}

It is interesting to perform the same check for the cross-section-level radiative 
gluon jet definition, which must reproduce the splitting kernel  
$P^{\mu\nu}_{g \rightarrow gg}$ when $m=1$. The diagrammatic contributions, in Feynman 
gauge, are similar to those in Fig. \ref{fig:cut_one_loop_jet}, and are displayed 
in Fig.~\ref{fig:cut_one_loop_jet_glue}; in an axial gauge, $n \cdot A = 0$, 
only the third graph, Fig.~(\ref{fig:cut_one_loop_jet_glue}\textcolor{blue}{c}), 
survives. In Feynman gauge, at amplitude level, the single-radiative jet function defined in 
\eq{collinear_current_glue} gives
\beq
  && \varepsilon^{* \, (\lambda)}_\mu (k) {\cal J}_{g, \, 1}^{\mu \nu, \, a} 
  \left(k_1; k, n \right) \, =  \, \frac{g_s \, t^{ a}}{(k_1 + k)^2} \,
  \left( \! - g^{\mu \nu} + \frac{n^\mu \, (k_1 + k)^\nu}{n \cdot (k_1 + k)} \right)
  \\ \nn && \hspace{2cm} \times 
  \bigg[ 2 \epsilon^*(k) \cdot  k_1 \, \epsilon^*_{\, \mu}(k_1)
  - 2 \epsilon^*(k_1) \cdot k \, \epsilon^*_{\, \mu}(k) 
  + \epsilon^*(k_1) \cdot \epsilon^*(k)\,  (k - k_1)_\mu \bigg] \; ,
\eeq
which can be verified to be consistent with the computation performed in axial 
gauge. Computing the single-radiative gluon jet function at cross-section level,
we can use the Sudakov parametrisation 
\beq
  k^\mu \, = \, z l^\mu +  l_\perp^\mu - \frac{l_\bot^2}{z} \frac{n^\mu}{2 l \cdot n} \, , 
  \qquad 
  k_1^\mu  \, = \, (1 - z) l^\mu  - l_\perp^\mu - \frac{l_\bot^2}{(1-z)} \frac{n^\mu}{2 l \cdot n} \, , 
\eeq
To leading power in $l_\bot$, and setting $n^2 = 0$, we end up with the expression
\beq
  && \sum_{\lambda_i} J_{g, \, 1}^{\mu \nu} \left( k; l, k_1, n \right) \, = \, 
  \frac{16 \pi \alpha_s C_A}{l^2} \, (2 \pi)^d \, \delta^d \left(l - k_1 - k \right) 
  \\ && \times \left[ - g^{\mu \nu} \left( \frac{z}{1 - z} + \frac{1 - z}{z} \right) - 
  2 \left(1 - \epsilon \right) z (1 - z) \, \frac{l_\bot^\mu l_\bot^\nu}{l_\bot^2}
   + \left( \frac{z}{1 - z} + \frac{1 - z}{z}\right) \frac{l^{ \{ \mu}n^{\nu \} }}{l \cdot n}
   \right] \, . \nn 
\eeq
The first two terms in the square bracket reproduce the expected splitting function;
the third term, where the braces denote index symmetrisation, is proportional to 
either $l^\mu$ or $l^\nu$: in the collinear limit, these corrections vanish when 
contracted with the factorised hard amplitude, which depends on the on-shell 
parent gluon momentum $l$. It is easy to check, by considering a final-state 
$q \bar{q}$ pair in \eq{collinear_current_glue}, that one may similarly recover 
the appropriate splitting function $P^{\mu\nu}_{g \rightarrow q \bar{q}}$; kernels
for double collinear emission can be reproduced with similar manipulations.

To complete our discussion, we note that the cross-section-level jet functions 
presented in \eq{qradjetsq} generate all collinear singularities, including soft-collinear 
ones. These are therefore double counted, since they were already included in the 
soft region. In order to avoid this issue, following the logic suggested by the 
factorisation of virtual corrections in \eq{factoramp}, we may introduce {\it radiative 
eikonal jet functions}, defined by replacing the field $\overline{\psi}(0)$ in 
\eq{qradjet} with a Wilson line (in the same colour representation), oriented 
along the hard parton direction $\beta^\nu = p^\nu/\mu$. At cross-section level, 
this leads to the definition
\beq
\label{eikjetsq}
  && J_{\E, \, m} \left(k_1, \ldots , k_m ; l, \beta, n \right) \, \equiv \,
  \sum_{p = 0}^\infty  J_{\E, \, m}^{(p)} \left(k_1, \ldots , k_m ; l, \beta, n \right)  \\
  && \equiv \, 
  \int d^d x \, {\rm e}^{{\rm i} l \cdot x} \, \bra{0}  
  \Phi_n (\infty, x) \Phi_\beta (x, \infty) \ket{k_j, \lambda_j}
  \bra{k_j, \lambda_j} \Phi_\beta (\infty, 0)
  \Phi_n (0, \infty) \ket{0} \nonumber \, .
\eeq
Notice that the radiative eikonal jet does not depend on the spin of the hard parton, 
so that \eq{eikjetsq} applies to gluons as well; the Fourier transform fixes $l^\mu$
to be the total momentum of the final state.

To test this definition, we compute the soft-collinear local counterterm for single 
radiation, and we easily find
\beq
  \sum_{ \lambda} J_{\E, \, 1} \left(k; l, \beta, n \right) \, = \, g_s^2 \, C_{\rm r} \,
  (2 \pi)^d \delta^d (l - p) \frac{2 p \cdot n}{p\cdot k \; n \cdot k} \, .
\label{eikjet_explic}
\eeq
In the limit of $p^\mu$ collinear to $k^\mu$, we can employ the relations 
\beq
  l^2 \, = \, ( p + k )^2 \, = \, 2 \, p \cdot k  \, , \qquad
  p \cdot n  \, = \, z \, l \cdot n \, , \qquad
  k \cdot n \, = \, (1 - z) \, l \cdot n \, ,
\eeq
to obtain the explicit soft-collinear counterterm
\beq
  \sum_{ \lambda} J_{\E, \, 1} \left(k; l, \beta, n \right) \, = \, 
  \frac{8 \pi \alpha_s C_{\rm r}}{l^2} \, (2 \pi)^d \delta^d (l - p) \, \frac{2 z}{1 - z} \, .
\label{eikjet_explic2}
\eeq
We note that the factor $2 z$ in the numerator is necessary to enforce the commutation
relation between soft and collinear limit at NLO: a basic feature that allows significant
simplifications in the subtraction procedure~\cite{Magnea:2018hab}.

%%%%%%%%%%%%%%%%%%%%%%%%%%%%%%%%%%%%%%%

\section{Constructing counterterms at NLO}
\label{NLOres}

Our basic strategy for subtraction is to identify soft and collinear local counterterms 
starting from the known expressions for the poles of virtual corrections. We now
proceed to illustrate how this works with the simple case of NLO massless final-states.
Expanding \eq{factoramp} to NLO, and using the fact that virtual
jet functions are normalised to equal unity at tree level, we easily find
\beq
  {\cal A}^{(0)}_n (p_i) & = &  {\cal S}^{(0)}_n (\beta_i) {\cal H}^{(0)}_n (p_i) \, , \nonumber \\
  {\cal A}^{(1)}_n (p_i) & = & 
  {\cal S}^{(1)}_n (\beta_i) {\cal H}^{(0)}_n (p_i) \, + \,
  {\cal S}^{(0)}_n (\beta_i) {\cal H}^{(1)}_n (p_i) \nonumber \\ && \qquad + \, 
  \sum_{i = 1}^n \left( {\cal J}^{(1)}_i (p_i) - {\cal J}_{i, \, \E}^{(1)} (\beta_i) \right) \,
  {\cal S}^{(0)}_n (\beta_i) \, {\cal H}^{(0)}_n (p_i)  \, ,
\label{oneloopamp}
\eeq 
Using \eq{oneloopamp}, it is straigthforward to construct the NLO virtual 
correction $V_n$, entering NLO distributions as in \eq{pertO}, and to express it 
in terms of the cross-section-level soft and jet virtual functions. One finds
\beq
  V_n & \equiv & 2 \, {\bf Re} \Big[ {\cal A}_n^{(0) *} {\cal A}_n^{(1)} \Big]  \\
  & = & {\cal H}^{(0) \, \dagger}_n (p_i) S_{n, \, 0}^{(1)} (\beta_i) {\cal H}^{(0)}_n (p_i)
  \, + \, \sum_{i = 1}^n \left( J^{(1)}_{i, \, 0} (p_i) - J_{i, \, \E, \, 0}^{(1)} (\beta_i) \right)
  \left| {\cal A}_n^{(0)} (p_i) \right|^2 \, + \, {\rm finite} \, . \nonumber
\label{virtol}
\eeq
It is now a simple task to find local counterterms for these poles: one simply notices
that the soft completeness relation in \eq{complete}, at NLO, implies the cancellation
\beq
  S_{n, 0}^{(1)} \, (\beta_i) \, + \, \int d \Phi_1 \, S_{n, \, 1}^{(0)} (k, \beta_i)
  \, = \, {\rm finite} \, .
\label{NLOsoftfin}
\eeq
Similarly, the collinear completeness relation in \eq{compcoll}, at NLO, implies the 
cancellation
\beq
  J_{i, \, 0}^{(1)} (l, p, n) \, + \, \int d \Phi_1 \, J_{i, \, 1}^{(0)} (k; l, p, n)
  \, = \, {\rm finite} \, ,
\label{NLOcollfin}
\eeq
with a similar relation holding for the cross-section-level eikonal jets defined in 
\eq{eikjetsq}. The local phase space integrands in \eq{NLOsoftfin} and \eq{NLOcollfin},
multiplied times the appropriate, finite, hard coefficients, must thus provide the
necessary counterterms. In particular NLO soft poles are cancelled by integrating
the combination
\beq
  K_{n+1}^{\, \rm{s}} \, = \, {\cal H}^{(0) \, \dagger}_n (p_i) \, 
  S_{n, \, 1}^{(0)} (k, \beta_i) \, {\cal H}^{(0)}_n (p_i) \, ,
\label{NLOsoftct}
\eeq
over the single-particle soft phase space. Similarly, NLO collinear poles are cancelled
by integrating the combination
\beq
  K_{n+1}^{\, \rm{c}} \, = \, \sum_{i = 1}^n J_{i, \, 1}^{(0)} (k_i; l, p_i, n_i) \, 
  \left| {\cal A}^{(0)}_n \left( p_1, \ldots, p_{i - 1}, l, p_{i +1}, \ldots, p_n \right) 
  \right|^2 \, ;
\label{NLOcollct}
\eeq
note that, for gluons, the function $J_{i, \, 1}$ is a spin matrix acting on the spin-correlated 
Born. The double subtraction of soft and collinear singularities overcounts the soft-collinear 
regions: one must therefore add back a local soft-collinear counterterm, given by
\beq
  K_{n+1}^{\, \rm{sc}} \, = \, \sum_{i = 1}^n J_{i, \, \E, \, 1}^{(0)} (k_i; l, \beta_i, n_i) \, 
  \left| {\cal A}^{(0)}_n \left( p_1, \ldots, p_{i - 1}, l, p_{i +1}, \ldots, p_n \right) 
  \right|^2 \, .
\label{NLOsoftcollct}
\eeq
Using the tree-level results listed in \secn{Softct} and in \secn {Collct}, it is easy 
to see that \eq{NLOsoftct} and \eq{NLOcollct} reproduce standard results for NLO
subtraction. One should however appreciate that the present approach provides
a simple proof that the list of singular regions for real radiation considered here
is exhaustive, and collinear regions for radiation from different outgoing hard 
particles do not interfere. While these facts are well-understood at NLO, their
generalisations at higher orders are much less obvious. On the other hand, 
we note that these result do not yet constitute a subtraction algorithm at NLO:
indeed, one can see that the tree-level matrix elements appearing in \eq{NLOcollct} 
involve particles that are not on the mass-shell, except in the strict collinear limit,
while momentum conservation is not properly implemented in \eq{NLOsoftct},
except in the strict soft limit. A practical algorithm must provide a resolution of 
these issues, with the construction of suitable momentum mappings between 
the Born and the radiative configurations, either with global treatment of phase
space, as done for example in \cite{Catani:1996vz}, or with a decomposition into 
different singular regions, as done for example in \cite{Frixione:1995ms} and 
in \cite{Magnea:2018hab}.

%%%%%%%%%%%%%%%%%%%%%%%%%%%%%%%%%%%%%%%

\section{Constructing counterterms at NNLO}
\label{NNLOres}

\begin{figure}[t]
 \centering
 {\includegraphics[scale=0.4]{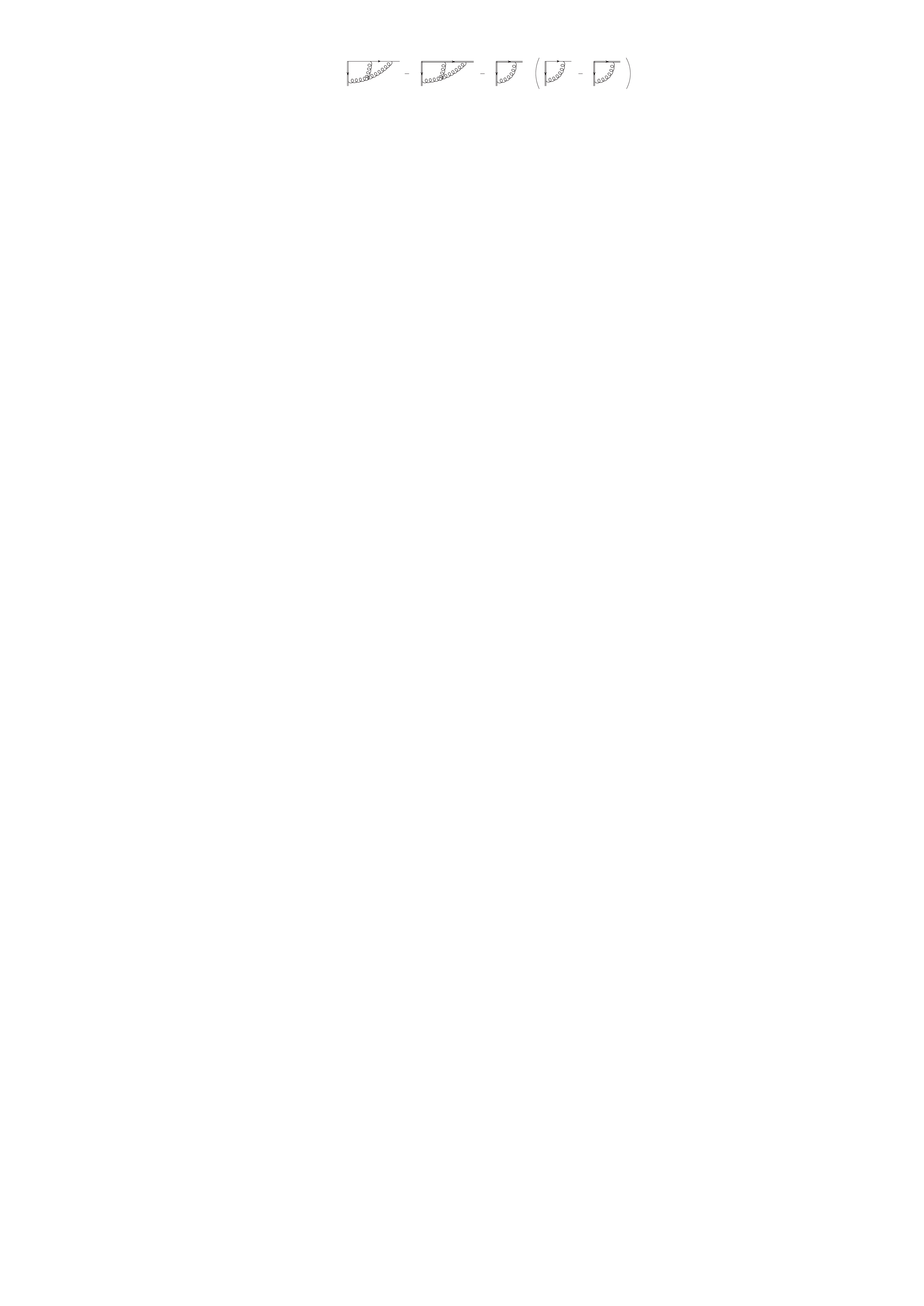}}
 \caption{Cancellation of soft poles illustrated with sample representative diagrams.}
 \label{fig:two_loop_jet_cancel}
 \end{figure}

Extending the procedure of \secn{NLOres} to higher orders is in principle straightforward,
but it unveils and organises several non-trivial features of real radiation in singular 
regions of phase space. Let us begin by extending \eq{oneloopamp} by computing
the expansion of the virtual correction to the amplitude up to NNLO. The two-loop
contributions can be written as
\beq
\label{twoloopamp}
  {\cal A}_n^{(2)} (p_i) & = &  {\cal S}^{(0)}_n (\beta_i) {\cal H}^{(2)}_n (p_i) \, + \, 
  {\cal S}^{(2)}_n (\beta_i) {\cal H}^{(0)}_n (p_i) \, + \, {\cal S}^{(1)}_n (\beta_i) 
  {\cal H}^{(1)}_n (p_i) \nonumber \\ 
  && + \,
  \sum_{i = 1}^n \bigg[ {\cal J}^{(2)}_i (p_i) - {\cal J}_{i, \, \E}^{(2)} (\beta_i) \, - \, 
  {\cal J}_{i, \, \E}^{(1)} (\beta_i) \left( {\cal J}^{(1)}_i (p_i) - 
  {\cal J}_{i, \, \E}^{(1)} (\beta_i) \right) \bigg] \,
  {\cal A}_n^{(0)} (p_i) \nonumber \\
  && + \, 
  \sum_{i < j = 1}^n \left( {\cal J}^{(1)}_i (p_i) - {\cal J}_{i, \, \E}^{(1)} (\beta_i) \right) 
  \left( {\cal J}^{(1)}_j (p_j) - {\cal J}_{j, \, \E}^{(1)} (\beta_j) \right) 
  {\cal A}_n^{(0)} (p_i) \\
  && + \, 
  \sum_{i = 1}^n \left( {\cal J}^{(1)}_i (p_i) - {\cal J}_{i, \, \E}^{(1)} (\beta_i) \right)
  \Big[ {\cal S}^{(1)}_n (\beta_i) {\cal H}^{(0)}_n (p_i) \, + \,
  {\cal S}^{(0)}_n (\beta_i) {\cal H}^{(1)}_n (p_i) \Big]  \, . \nonumber 
\eeq 
Several comments are in order. We begin by noting that the first term on the first line is 
finite, being given by the action of the finite tree-level soft operator on the two-loop
finite hard remainder. The second term contains two-loop soft and soft-collinear
poles from the soft operator, giving singularities up to the maximum allowed
degree, $1/\eps^4$. In the third term the one-loop soft operator acts on the 
one-loop finite hard remainder, giving a single soft pole and a double soft-collinear 
pole. The second line is the most interesting from the point of view of factorisation:
it contains all double hard-collinear poles arising from two-loop virtual corrections 
associated with a single hard external leg, yielding singularities up to $1/\eps^2$.
In particular, the second line does not generate any soft poles: indeed, while 
the function  ${\cal J}^{(2)}_i (p_i)$ contains up to two soft poles, generated by 
gluons that are both soft and collinear to the $i$-th hard particle, the contributions
in which both gluons are soft (on top of being collinear) are cancelled by the
second term in square bracket, ${\cal J}_{i, \, \E}^{(2)} (\beta_i)$, and finally 
the contributions in which only one of the two collinear gluons is soft are 
cancelled by the last term in the square bracket. Notice the factorised form 
of that last term: when one gluon is hard and the other one is soft, the soft
gluon factorises from the matrix element in the usual way. This cancellation
mechanism is illustrated, for a sample diagram, in Fig.~\ref{fig:two_loop_jet_cancel}.
The last two lines in \eq{twoloopamp} have a simpler interpretation:
the third line contains single hard collinear poles arising simultaneously
on two different hard legs, $i$ and $j$; the fourth line contains single hard collinear 
poles on the $i$-th hard leg, accompanied by a soft single pole, or a soft-collinear
double pole, or just multiplied times a finite correction.

The next step is to construct the virtual contributions to the squared amplitude 
at NNLO. In order for our procedure to work, these must in turn be expressed
in terms of the cross-section-level virtual jet and soft functions, which is less 
than trivial since, at NNLO, all functions involved receive contributions both
from the interference between the Born amplitude and the two-loop correction,
and from the square of the one-loop amplitudes. For example, the two-loop
cross-section-level virtual soft function is given by
\beq
  S_n^{(2)} \, = \, {\cal S}_n^{(0) \, \dagger} {\cal S}_n^{(2)} \, + \, 
  {\cal S}_n^{(2) \, \dagger} {\cal S}_n^{(0)} \, + \,  {\cal S}_n^{(1) \, \dagger} 
  {\cal S}_n^{(1)} \, ,
\label{sigmasoft}
\eeq
while the two-loop unpolarised cross-section-level radiative jet function for 
a quark emitting $m$ gluons reads
\beq
   J_{q, \, m}^{(2)} & = & \int d^d x \, {\rm e}^{{\rm i} l \cdot x} \sum_{\{\lambda_j\}}
   \bigg[ {\cal J}_{q, \, m}^{(1) \, \dagger} (x) \, \slash{p} \, {\cal J}_{q, \, m}^{(1)} (0)   
   \, + \, {\cal J}_{q, \, m}^{(0) \, \dagger} (x) \, \slash{p} \, {\cal J}_{q, \, m}^{(2)} (0)
   \nonumber \\  && \hspace{3cm} 
   + \, {\cal J}_{q, \, m}^{(0)} (x) \, \slash{p} \, {\cal J}_{q, \, m}^{(2) \, \dagger} (0)
   \bigg] \, .
\label{sigmajet}
\eeq
It is relatively simple to organise the virtual poles in the real-virtual contribution to
the squared matrix element: this amounts essentially to a repetition of the NLO
calculation, with $n+1$ hard particles in the final state. One easily finds
\beq
\label{rvirtol}
  && RV_\npo \, \equiv \, 2 \, {\bf Re} \Big[ {\cal A}_\npo^{(0) *} 
  {\cal A}_\npo^{(1)} \Big] \\
  && \hspace{-1cm} = \, {\cal H}^{(0) \, \dagger}_\npo \, S_{\npo, \, 0}^{(1)} \, 
  {\cal H}^{(0)}_{n + 1} \, + \, 
  \sum_{i = 1}^\npo \left( J^{(1)}_{i, \, 0} - J_{i, \, \E, \, 0}^{(1)} \right)
  \left| {\cal A}_\npo^{(0)} \right|^2 \, + \, {\rm finite} \, . \nonumber 
\eeq
Double virtual poles, on the other hand, receive several non-trivial contributions, 
which can be organised as follows:
\beq
\label{vvirtol}
  VV_{n} & \equiv &  \left( VV \right)_{n}^{(2 {\rm s})} \, + \,  
  \left( VV \right)_{n}^{(1 {\rm s})} \, + \, \sum_{i = 1}^n 
  \left( VV \right)_{n, \, i}^{(2 {\rm hc})} \, + \,
  \sum_{i < j = 1}^n \left( VV \right)_{n, \, ij}^{(2 {\rm hc})} 
  \nonumber \\  && 
  + \, \sum_{i = 1}^n \left( VV \right)_{n, \, i}^{(1 {\rm hc},\,  1 {\rm s})} \, + \,
  \sum_{i = 1}^n \left( VV \right)_{n, \, i}^{(1 {\rm hc})}  \, .
\eeq
We will now go through the various contributions to the {\it r.h.s.} of \eq{vvirtol}, 
identifying in each case the real radiation counterterms that are needed to cancel 
the corresponding virtual poles. The double-soft virtual contribution $(VV)_{n}^{(2 
{\rm s})}$, which, within our chosen organisation of the matrix element, contains 
soft-collinear poles as well, is given by
\beq
  \left( VV \right)_{n}^{(2 {\rm s})} 
  \, = \,  {\cal H}^{(0) \,  \dagger}_n \, S_{n, \, 0}^{(2)} \, {\cal H}^{(0)}_n \, ,
\label{vvirtolSS} 
\eeq
where $S_{n, \, 0}^{(2)}$ was given in \eq{sigmasoft}.
To give a complete picture of the soft sector, at this point we also include in the 
discussion the single-soft virtual contribution $(VV)_{n}^{(1 
{\rm s})}$, which is given by
\beq
  \left( VV \right)_{n}^{(1 {\rm s})} 
  \, = \, {\cal H}^{(0) \,  \dagger}_n S_{n, \, 0}^{(1)} {\cal H}^{(1)}_n \, + \,
  {\cal H}^{(1) \,  \dagger}_n S_{n, \, 0}^{(1)} {\cal H}^{(0)}_n  \, ,
\label{vvirtolS} 
\eeq
as well as the real-virtual soft poles in \eq{rvirtol}. In order to cancel these poles, we 
need the completeness relation for the soft sector to NNLO, which reads
\beq
  S_{n, 0}^{(2)} (\beta_i) \, + \, \int d \Phi_1 \, S_{n, \, 1}^{(1)} (k, \beta_i) \, + \, 
  \int d \Phi_2 \, S_{n, \, 2}^{(0)} (k_1, k_2, \beta_i) \, = \, {\rm finite} \, .
\label{NNLO2softfin}
\eeq
It is natural at this point to identify three separate soft counterterms, characterised 
by their kinematic structure. The double-unresolved soft counterterm involves
$n$-point hard kinematics, and double soft radiation; it is given by
\beq
  K^{(2 {\rm s})}_\npt & = & 
  {\cal H}^{(0) \,  \dagger}_n \, S_{n, \, 2}^{(0)} \, {\cal H}^{(0)}_n \, .
\label{NNLOsoftcts_ss}
\eeq
The single-unresolved soft conterterm involves $(n+1)$-point hard kinematics, and
single soft radiation; it is given by
\beq
  K^{({\bf 1}, \, {\rm s})}_\npt & = & 
  {\cal H}^{(0) \,  \dagger}_\npo \, S_{\npo, \, 1}^{(0)} \, {\cal H}^{(0)}_\npo \, .
\label{NNLOsoftcts_s}
\eeq
Finally, the real-virtual soft counterterm involves $n$-point hard kinematics, and
single soft radiation; it contains all remaining terms that are required for finiteness
according to Eqs.~(\ref{NLOsoftfin}) and (\ref{NNLO2softfin}), which give
\beq
  K^{({\bf RV}, \, {\rm s})}_\npo & = & 
  {\cal H}^{(0) \,  \dagger}_n \, S_{n, \, 1}^{(0)} \, {\cal H}^{(1)}_n \, + \, 
  {\cal H}^{(1) \,  \dagger}_n \, S_{n, \, 1}^{(0)} \, {\cal H}^{(0)}_n \, + \,
  {\cal H}^{(0) \,  \dagger}_n \, S_{n, \, 1}^{(1)} \, {\cal H}^{(0)}_n  \, .
\label{NNLOsoftcts_rv}
\eeq
We now note that this procedure yields an expression for the complete double-unresolved
soft counterterm $K^{(2 {\rm s})}_\npt$, but does not immediately distinguish between 
the two contributions defined in \eq{splitdouble}. It is however easy, in this context, 
to identify the desired partition of the counterterm. Indeed, as discussed in 
Ref.~\cite{Magnea:2018hab}, the local counterterm $K^{\otwo}_\npt$ is designed 
to be integrated in two stages: the first integration, in a single-particle phase space, 
yields the integrated counterterm $I^{\otwo}_\npo$, which must cancel the explicit 
poles of the real-virtual counterterm $K^{\RV}_\npo$, given entirely by the last 
term in \eq{NNLOsoftcts_rv}. From the point of view of factorisation, the desired 
function is then identified as follows. An explicit calculation of $S_{n, \, 2}^{(0)}$ 
from its definition in \eq{softsigma} yields the sum of two distinct contributions: one
in which the soft limits on the two radiated gluons are taken hierarchically, with one 
gluon being much softer than the other one, and one in which the two gluons have
a comparable softness. This structure was identified in Ref.~\cite{Catani:1999ss},
and is derived from \eq{treesoftcurfull} by taking the limit in which $k_2$ is much 
softer than $k_1$, or viceversa. The hierarchical limit of $K^{(2 {\rm s})}_\npt$ is 
constructed essentially by treating one of the two soft radiated particles temporarily 
as a hard one: it gives therefore precisely the desired function $K^{({\bf 12},\, 2\rm s)}_\npt$, 
which, upon integration, will cancel the explicit double-soft poles of the real-virtual local counterterm.
A similar pattern can be replicated for the other double-unresolved local counterterms,
in all cases in which a hierarchy between the two unresolved particles can be 
identified.

Turning to hard collinear poles, we first tackle the contribution with two hard collinear
virtual gluons attached to the same hard outgoing leg. It is given by
\beq
  \left( VV \right)_{n, \, i}^{(2 {\rm hc})} 
  \, = \, \bigg[ J^{(2)}_{i, \, 0} - J_{i, \, \E, \, 0}^{(2)} \, - \, 
  J_{i, \, \E, \, 0}^{(1)} \left( J^{(1)}_{i, \, 0} - J_{i, \, \E, \, 0}^{(1)} \right) \bigg] \, 
  \left| {\cal A}_n^{(0)} \right|^2 \, .
\label{vvirtol2hci}
\eeq
In order to cancel the poles of the first two terms in \eq{vvirtol2hci}, we can use 
the NNLO expansion of \eq{compcoll}, which gives the finiteness condition
\beq
  J_{i, 0}^{(2)} \, + \, \int d \Phi_1 \, J_{i, \, 1}^{(1)} \, + \, 
  \int d \Phi_2 \, J_{i, \, 2}^{(0)} \, = \, {\rm finite} \, ,
\label{NNLO2collfin}
\eeq
and the analogous expression for eikonal jets.
The third term of \eq{vvirtol2hci} has a different structure, since it is a product of two
one-loop functions. One can however cancel its poles with the same general
approach, by using the fact that
\beq
  \hspace{-4mm}
  \left[ J_{i, \, \E, 0}^{(1)} \, + \, \int d \Phi_1 \, J_{i, \, \E, \, 1}^{(0)} \right] \left[ 
  J_{i, 0}^{(1)} \, - \, J_{i, \, \E, 0}^{(1)} \, +  \int d \Phi_1' \left( J_{i, \, 1}^{(0)} \, - \, 
  J_{i, \, \E, 1}^{(0)} \right) \right] \, = \, {\rm finite} \, .
\label{NNLO2collfin2}
\eeq
Once again, the contributions to different local counterterm functions can be identified
by their phase space structure. We define
\beq
  K^{(2{\rm hc})}_{\npt, \, i} & = & 
  \bigg[ J^{(0)}_{i, \, 2} - J_{i, \, \E, \, 2}^{(0)} \, - \, 
  J_{i, \, \E, \, 1}^{(0)} \left( J^{(0)}_{i, \, 1} - J_{i, \, \E, \, 1}^{(0)} \right) \bigg]
  \left| {\cal A}^{(0)}_n \right|^2 \, , \nonumber \\
  K^{({\bf 1}, \, {\rm hc})}_{\npt, \, i} & = & 
  \left( J^{(0)}_{i, \, 1} - J_{i, \, \E, \, 1}^{(0)} \right) \left| {\cal A}^{(0)}_\npo \right|^2 \, , \\
  K^{({\bf RV}, \, {\rm hc})}_{\npo, \, i} & = & 
  \left[ J^{(1)}_{i, \, 1} - J_{i, \, \E, \, 1}^{(1)} - 
  J_{i, \, 0}^{(1)} J^{(0)}_{i, \, \E, \, 1} - J_{i, \, \E, \, 0}^{(1)} J^{(0)}_{i, \, 1} +
  2 J_{i, \, \E, \, 0}^{(1)} J^{(0)}_{i, \, \E, \, 1}
  \right] \left| {\cal A}^{(0)}_n \right|^2 \, . \nonumber
\label{NNLOcollctsi}
\eeq
The remaining singular virtual contibutions do not present new difficulties. Hard collinear
virtual poles associated with two different hard legs can be organised in the form
\beq
  \left( VV \right)_{n, \, ij}^{(2 {\rm hc})} & = & \left( J^{(1)}_{i, \, 0} - 
  J_{i, \, \E, \, 0}^{(1)} \right) \left( J^{(1)}_{j, \, 0} - J_{j, \, \E, \, 0}^{(1)} \right)  \, 
  \left| {\cal A}_n^{(0)} \right|^2 \, .
\label{vvirtol2hcij}
\eeq
By using again the finiteness conditions stemming from \eq{compcoll} (and its eikonal 
counterpart), we can cancel these poles by integrating the local counterterms
\beq
  K^{(2 {\rm hc})}_{\npt, \, ij} & = & 
  \left( J^{(0)}_{i, \, 1} - J_{i, \, \E, \, 1}^{(0)} \right) 
  \left( J^{(0)}_{j, \, 1} - J_{j, \, \E, \, 1}^{(0)} \right) 
  \left| {\cal A}^{(0)}_n \right|^2 \nonumber \\
  K^{({\bf RV}, \, {\rm hc})}_{\npo, \, ij} & = & 
  \left[ \left( J^{(1)}_{i, \, 0} - J_{i, \, \E, \, 0}^{(1)} \right)
  \left( J^{(0)}_{j, \, 1} - J_{j, \, \E, \, 1}^{(0)} \right) \, + \, 
  \left( i \leftrightarrow j \right) \right] 
  \left| {\cal A}^{(0)}_n \right|^2 \, ,
\label{NNLOcollctsij}
\eeq
while no single-unresolved counterterm in the $(\npo)$-particle phase space is required
in this case.

We are left with single hard collinear virtual poles, accompanied by a single soft pole,
or by a finite factor. They are given by
\beq
  \left( VV \right)_{n, \, i}^{({1\rm hc} ,\,  {1\rm s})} & = & \left( J^{(1)}_{i, \, 0} - 
  J_{i, \, \E, \, 0}^{(1)} \right) \, 
  {\cal H}^{(0) \,  \dagger}_n S_{n, \, 0}^{(1)} {\cal H}^{(0)}_n \, , 
  \label{vvirtolrest} \\
  \left( VV \right)_{n, \, i}^{(1{\rm hc})} & = & \left( J^{(1)}_{i, \, 0} - 
  J_{i, \, \E, \, 0}^{(1)} \right) \, \left(
  {\cal H}^{(0) \,  \dagger}_n S_{n, \, 0}^{(0)} {\cal H}^{(1)}_n \, + \,
  {\cal H}^{(1) \,  \dagger}_n S_{n, \, 0}^{(0)} {\cal H}^{(0)}_n \right) \, . \nonumber
\eeq
Proceeding as above, we find that these poles can be cancelled by integrating
the local counterterms
\beq
  K^{(1{\rm hc} , \, {1\rm s})}_{\npt, \, i} & = & 
  \left( J^{(0)}_{i, \, 1} - J_{i, \, \E, \, 1}^{(0)} \right) 
  {\cal H}^{(0) \,  \dagger}_n S_{n, \, 1}^{(0)} {\cal H}^{(0)}_n 
  \, , \label{NNLOrest} \\
  K^{({\bf RV}, \, {1\rm hc} , \,  {1\rm s})}_{\npo, \, i} & = & 
  \left( J^{(1)}_{i, \, 0} - J_{i, \, \E, \, 0}^{(1)} \right)
  {\cal H}^{(0) \,  \dagger}_n S_{n, \, 1}^{(0)} {\cal H}^{(0)}_n \, + \, 
  \left( J^{(0)}_{i, \, 1} - J_{i, \, \E, \, 1}^{(0)} \right)
  {\cal H}^{(0) \,  \dagger}_n S_{n, \, 0}^{(1)} {\cal H}^{(0)}_n \, , \nonumber \\
  K^{({\bf RV}, \, {1\rm hc})}_{\npo, \, i} & = & 
  \left( J^{(0)}_{i, \, 1} - J_{i, \, \E, \, 1}^{(0)} \right)
  \left( {\cal H}^{(0) \,  \dagger}_n S_{n, \, 0}^{(0)} {\cal H}^{(1)}_n \, + \,
  {\cal H}^{(1) \,  \dagger}_n S_{n, \, 0}^{(0)} {\cal H}^{(0)}_n \right) \, ,
  \nonumber
\eeq
which completes the list of local counterterms needed for NNLO massless
final state configurations.

%%%%%%%%%%%%%%%%%%%%%%%%%%%%%%%%%%%%%%%

\section{Conclusions}
\label{Conclu}

We have presented the outline of a general formalism to construct local counterterms 
for the subtraction of soft and collinear singular configurations from real-radiation
squared matrix elements, using as an input the well-known factorised structure
of infrared poles in virtual corrections to scattering amplitudes. Virtual factorisation
embodies highly non-trivial structural features of infrared singularities: the colour-singlet 
nature of collinear poles, the simple organisation of soft-collinear enhancements,
the exponentiation of singularities following from renormalisation group invariance.
The hope, already  partly realised in the results presented here, is that these
simplifying features will be reflected in a streamlined and optimised structure
of the subtraction procedure.

The main results of this paper are presented in \secn{Softct} and in \secn{Collct},
where we give general expressions for local counterterms for soft, collinear and 
soft-collinear configurations, valid to all orders in perturbation theory, and constructed
in terms of gauge-invariant matrix elements of field operators and Wilson lines.
The definitions are tested at low orders, reproducing known results at NLO and 
NNLO and highlighting the simplifying features that follow from virtual factorisation.
In \secn{NLOres} and in \secn{NNLOres} we apply the general definitions to
construct explicitly all counterterms required at NLO and at NNLO, respectively,
for the case of massless final state radiation. We emphasise that the expressions
given here are not yet directly suitable for implementation in a fully operational 
subtraction algorithm: appropriate phase-space mappings, such as those detailed 
in~\cite{Magnea:2018hab}, must still be implemented in order to express all 
ingredients in terms of on-shell momentum-conserving matrix elements; we 
note however that the list of counterterms presented is exhaustive, and the 
treatment of soft-collinear double counting is highly streamlined.

The approach we have presented can be naturally generalised in several 
directions: first of all, a detailed treatment of initial-state singularities can be 
developed, which in principle does not present new theoretical difficulties. In 
this context, we note that we are not paying special attention to the issue of
Glauber gluons and possible factorisation violations: essentially, we are 
assuming that the formalism applies for sufficiently inclusive observables, 
such that Glauber gluons do not result in uncancelled infrared singularities.
The issue is however very interesting from a theoretical point of view: infrared
power counting shows that Glauber gluons do not contribute to infrared 
singularities for fixed-angle scattering amplitudes (see, for example, 
\cite{Erdogan:2014gha} for a recent discussion of leading regions in coordinate 
space), but they can result in a breakdown of factorisation for insufficiently 
inclusive hadronic cross sections, when real collinear radiation is integrated
over unresolved regions of phase space (see, for example, \cite{Aybat:2008ct,
Catani:2011st,Rothstein:2016bsq} for recent discussions). The tools developed 
in this paper, which allow for the study of real infrared radiation at the level 
of differential distributions, may in future help to shed light on the limits of 
factorisation theorems for less inclusive collider observables.

At the level of definitions of local counterterms, the extension to massive partons
(which is of obvious phenomenological interest, in view of top-quark-related
observables, and possibly $b$-quark mass effects) is not problematic: indeed,
massive partons are not affected by collinear divergences (although it may
be of interest to resum collinear logarithms of the quark mass), so that 
the structure of counterterms is in fact simpler when masses are present. 
In the massive case, on the other hand, more work is needed to properly
define the phase space mappings associated with branchings involving 
massive partons~\cite{Catani:2002hc}, and to perform the corresponding integrations.
On the other hand, the approach
we have presented here is likely to have a significant impact in the organisation 
of future N$^3$LO subtraction algorithms: indeed, at N$^3$LO, the combinatorics
of overlapping singular regions becomes considerably worse, and the impact
of infrared exponentiation on subtraction is bound to become stronger. Work
on a detailed extension of the present work to N$^3$LO is in progress.

More generally, we hope that the present work will contribute to developing
our knowledge of the infrared behaviour of real radiation at the differential 
level, to all orders in perturbation theory, bringing it to the same detailed level
of understanding and control currently enjoyed by virtual corrections to fixed-angle
scattering amplitudes and by inclusive cross sections.

%%%%%%%%%%%%%%%%%%%%%%%%%%%%%%%%%%%%%%%

\section*{Acknowledgments}

The work of PT has received funding from the European Union Seventh 
Framework programme for research and innovation under the Marie Curie 
grant agreement N. 609402-2020 researchers: Train to Move (T2M).

%%%%%%%%%%%%%%%%%%%%%%%%%%%%%%%%%%%%%%%

%\appendix

%%%%%%%%%%%%%%%%%%%%%%%%%%%%%%%%%%%%%%%

\bibliographystyle{JHEP}
\bibliography{fact}

%%%%%%%%%%%%%%%%%%%%%%%%%%%%%%%%%%%%%%%

\end{document}